\documentclass[superscriptaddress,groupedaddress,nofootnoteinbib,12pt,english]{article}
 \usepackage[latin9]{inputenc}
\usepackage[letterpaper]{geometry}
\usepackage{amsmath}
\usepackage{amssymb}
\usepackage{cite}
\usepackage{babel}
\usepackage{jcappub}

\makeatletter

\newcommand{\lag}{\mathcal{L}}

\newcommand{\la}{\langle}
\newcommand{\ra}{\rangle}

\newcommand{\Lm}{\Lambda}
\newcommand{\ep}{\epsilon}

\newcommand{\intdx}{\int\rmd ^{4}x}
\newcommand{\rmd}{{\rm d}}
\newcommand{\bx}{\square}
\newcommand{\pd}{\partial}
\newcommand{\pdt}{\partial_{t}}
\newcommand{\pdr}{\partial_{r}}
\newcommand{\pdO}{\nabla_{\Omega}}

\newcommand{\Mom}{\mathcal{M}}

\newcommand{\Mpl}{M_{{\rm Pl}}}
\newcommand{\Mtot}{M_{{\rm tot}}}
\newcommand{\Mred}{\mathcal{M}}

\newcommand{\rstar}{r_{\star}}
\newcommand{\rstarThree}{r_{\star,3}}
\newcommand{\rstarFour}{r_{\star,4}}

\newcommand{\lcrit}{\ell_{{\rm crit}}}

\newcommand{\Op}{\Omega_{P}}
\newcommand{\Tp}{T_{P}}
\newcommand{\rbar}{\bar{r}}
\newcommand{\nmax}{n_{{\rm max}}}

\newcommand{\phiFirstOrder}{\phi^{(1)}}
\newcommand{\phiSecondOrder}{\phi^{(2)}}

\def\be{\begin{equation}}
\def\ee{\end{equation}}

\def\ba{\begin{eqnarray}}
\def\ea{\end{eqnarray}}

\def\nn{\nonumber}

\def\d{\mathrm{d}}

\def\({\left(}
\def\){\right)}

\def\ie{{\it i.e. }}

\def\Op{\Omega_P}
\def\l{\ell}

\def\mpl{M_{\rm Pl}}
\def\p{\partial}
\def\stu{St\"uckelberg }

\makeatother

\begin{document}

\title{Galileon Radiation from Binary Systems}
\author{Claudia de Rham, Andrew Matas and Andrew J. Tolley}

\affiliation{Department of Physics, Case Western Reserve University, 10900 Euclid Ave, Cleveland, OH 44106, USA}

\abstract{
We calculate the power emitted in scalar modes for a binary systems, including binary pulsars,
with a conformal coupling to the most general Galileon effective field
theory by considering perturbations around a static, spherical background.
While this method is effective for calculating the power in the cubic
Galileon case, here we find that if the quartic or quintic Galileon dominate, for realistic pulsar systems the classical perturbative expansion about spherically symmetric backgrounds breaks down (although the quantum effective theory is well-defined).
The basic reason is that the equations of motion for the fluctuations
are then effectively one dimensional. This leads to many multipoles radiating
with equal strength, as opposed to the normal Minkowski spacetime and cubic
Galileon cases, where increasing multipoles are suppressed by increasing
powers of the orbital velocity. We consider two cases where perturbation
theory gives trust-worthy results: (1) when there is a large hierarchy
between the masses of two orbiting objects, and (2) when we choose
scales such that the quartic Galileon only begins to dominate at distances
smaller than the inverse pulsar frequency. Implications for future
calculations with the full Galileon that account for the Vainshtein
mechanism are considered.
}

\maketitle

\section{Introduction}

Attempts to explain the observation of a small but nonzero vacuum
energy in a dynamical way generically introduce new light
scalar degrees of freedom, \cite{Khoury:2010xi,deRham:2012az}.
Frequently these scalars are themselves radiatively unstable.
For example, in massive gravity, the new
helicity-0 polarization state of the graviton acts as a light scalar.
As such light degrees of freedom have not been detected in precision
tests of gravity, the phenomenological viability of these theories
rests on the existence of screening mechanisms that hide the scalars
on small scales where gravity has been well tested. Therefore it is
imperative to understand in detail how screening mechanisms work.
In this work we will study how one such screening mechanism, the Vainshtein
mechanism \cite{Vainshtein:1972sx}, operates around dynamical,
non-relativistic sources. Vainshtein
screening works by making the scalars strongly coupled to themselves
near compact sources, essentially suppressing their coupling to matter.

While studying the Dvali-Gabadadze-Porrati (DGP) model of `soft' massive gravity \cite{Dvali:2000rv}, a class of
scalar field models was discovered that exhibit the Vainshtein mechanism.
These fields posses the galilean symmetry $\pi\rightarrow\pi+c+v_{\mu}x^{\mu}$
and are thus referred to as `Galileons' \cite{Nicolis:2008in}. They arise naturally in the
decoupling limit of theories of massive gravity, both `soft' such
as DGP \cite{Nicolis:2004qq}, and `hard' such as massive gravity from auxiliary extra dimensions, \cite{Gabadadze:2009ja,deRham:2009rm,deRham:2010gu}, ghost free dRGT massive gravity \cite{deRham:2010ik,deRham:2010kj}, or New Massive Gravity in three dimensions, \cite{Bergshoeff:2009hq,deRham:2011ca}.
They can also be considered as an effective field theory in their own right,
which is the approach we will take in this work.
Vainshtein screening for Galileons around static, spherically symmetric
is well understood and is quite effective \cite{Nicolis:2004qq,Nicolis:2008in}.
There has recently been a lot of work studying the Vainshtein mechanism
\cite{Belikov:2012xp,
Hiramatsu:2012xj,
Sbisa:2012zk,
Iorio:2012pv,
Hui:2012jb,
Kimura:2011dc,
Sjors:2011iv,
DeFelice:2011th, 
Babichev:2011iz,
Babichev:2011kq,
Koyama:2011yg,
Koyama:2011xz,
Gannouji:2011qz, 
Kaloper:2011qc, 
Chkareuli:2011te,
Wyman:2011mp,
Hui:2010dn,
Babichev:2010jd,
Babichev:2009jt,
Hui:2009kc,
Babichev:2009us}.

However, recently it has
been established that within its simplest realization (\ie in the cubic Galileon),  the Vainshtein mechanism is slightly less effective around dynamic sources, because the orbital period introduces a new large
length scale $\Op ^{-1}$ \cite{deRham:2012fw}. There it was shown that the Vainshtein
suppression to the gravitational radiation is $(\Op \rstar )^{-3/2}$,
instead of the naive expectation from static sources $(\rbar /\rstar )^{3/2}$,
where $\rbar $ is the size of the system and $\rstar $ is the Vainshtein
radius.
 Nevertheless, the Vainshtein mechanism in that case ends up being still powerful enough to evade any constraints from pulsar systems.
A similar problem was also recently considered in \cite{Chu:2012kz}.

In this work we extend the results to the power emitted by binary
pulsars in the presence of all four non-trivial Galileon terms possible
in four dimensions. We consider a conformal $\pi T$ coupling to matter.
We find that, surprisingly, the naive perturbation method used for instance in \cite{Goldberger:2004jt} fails
to work in this case. This is despite the fact that the perturbative expansion is valid for the cubic Galileon case \cite{deRham:2012fw} as we show explicitly here in Appendix \ref{sec:App5}.
The difference is that when the quartic Galileon dominates, the effective one dimensional metric seen by
the fluctuations means modes radiate with an effective frequency
\begin{equation}\label{eq:omega_l}
\omega_{\ell}^{2}=\omega^{2}-\frac{\ell(\ell+1)}{\rstar ^{2}}\,.
\end{equation}
Multipoles radiate as long as their associated effective frequency $\omega_{\ell}$ is real. The crucial point is that there is no distinction
between modes of different $\ell$ when $\ell\ll\omega \rstar $, so
each multipole radiate with equal strength. Since $\Op \rstar \gg1$
(for the Hulse-Taylor pulsar $\Op \rstar \sim10^{6}$, when considering a strong coupling scale $\Lambda\sim (1000{\rm km})^{-1}$), a huge number of modes radiate, each with a comparable power.
Even more multipoles radiate when we consider the power emitted by higher harmonics
with frequency $n\Op $ for integer $n$.

We interpret this result as a failure of perturbation theory, and
explicitly demonstrate that the perturbation series breaks down. While
we are able to find some regimes where perturbation theory is valid
and obtain sensible results, the mere fact that perturbation theory breaks
down  in this situation  questions the intuition one can really infer from
the static Vainshtein mechanism when dealing with more complicated
systems.

We stress that this breakdown of perturbation theory is not a quantum strong coupling problem. The quantum low energy effective theory for the Galileon remains under control, rather this is the breakdown of the description of the classical field configuration for the Galileon around the pulsar as a nonlinear spherical profile plus small time-dependent non-spherical perturbations. It is thus a failure of the approximation method used to calculate radiation and not of the effective field theory per se.

These results suggest that there is additional Vainshtein suppression on top of the
naive expectation from the static, spherically symmetric case.
We obtain an estimate for the power by introducing a cutoff  $\ell$ in the number of multipoles
that we sum to determine the power. Physically this is motivated by the fact that
the higher order multipoles are more strongly angular dependent and so
more sensitive to the non-spherical nature of the system. We expect that
multipoles at higher $\ell$ will become nonlinear and see Vainshtein screening
on top of the Vainshtein screening from the background. With this procedure
we find that the total power obtained is more Vainshtein suppressed $\sim \rstar^{-3}$
than any one mode $\sim \rstar^{-2}$.

The rest of this paper as organized as follows. In Section \ref{sec:Pert} we derive the
perturbation equations around a static, spherically symmetric background and review how to use the Feynman
propagator to derive the power emitted. In Section \ref{sec:Power_1} we apply the
naive perturbative methods of \cite{Goldberger:2004jt,deRham:2012fw}
to compute the power in a binary pulsar system, and find that in that case the
resulting power is formally divergent. We then check whether
perturbation theory is under control by explicitly constructing the first and
second order solutions using the retarded Green function and comparing them.
We find that, unlike the cubic Galileon case \cite{deRham:2012fw} the orbital velocity small
parameter $v$ is not enough to ensure the convergence of the perturbation series, and one needs an
additional small parameter to have a controlled expansion around a spherically symmetric background.
This can take the form of a hierarchy between
the two masses in the binary system, or between the strong coupling scale
for the cubic Galileon and the strong coupling scale for the quartic Galileon.
We consider these two cases in sections \ref{sec:MassHierarchy} and \ref{sec:SC_hierarchy} respectively
as a check that the Galileon theory is well behaved in regimes where we can trust perturbation theory.
We conclude by considering possible methods to calculate the power emission
and the implications of this result for other studies of the Vainshtein mechanism.

\section{Perturbations in a Galileon Theory}
\label{sec:Pert}

We consider a Galileon theory around Minkowski spacetime. These theories can arise from fully
covariant theories in a number of different ways \cite{Deffayet:2009wt,deRham:2010eu}.
For example one can start with the action for ghost-free massive gravity \cite{deRham:2010kj}
\be\label{eq:massive-gravity-action}
S=\frac{\Mpl^2}{2}\intdx \sqrt{-g}\(R-\frac{m^2}{4}\mathcal{U}(g,H)\)\,,
\ee
where the tensor $H_{\mu\nu}$  is defined in terms of the metric and the four \stu fields $\phi^a$  as
\be\label{eq:def-H}
g_{\mu\nu}=\pd_\mu \phi^a \pd_\nu \phi^b \eta_{ab}+H_{\mu\nu}\,.
\ee
The ghost-free potential $\mathcal{U}(g,H)$ is fixed up to two free parameters (in addition to the mass parameter $m$).
The Galileon effective field theory then arises in the decoupling limit $\Mpl\rightarrow\infty,\ m\rightarrow 0$ keeping the scale $\Lambda= \(m^2 \Mpl\)^{1/3}$ fixed, where the helicity-0 mode plays the role of the Galileon.

Regardless of the full covariant completion, we start with a Galileon scalar field $\pi$ in Minkowski spacetime
conformally coupled to matter \cite{Nicolis:2008in} and consider all possible interactions that respect the Galileon properties in four dimensions (ignoring the tadpole),
\begin{equation}\label{eq:dec-action}
S=\intdx \left(-\frac{1}{4}h^{\mu\nu}(\mathcal{E}h)_{\mu\nu}-\frac{3}{4}\sum_{i=2}^{5}\frac{c_i \lag_{i}}{\Lm_{i}^{3(i-2)}}
+\frac{1}{2\Mpl}h^{\mu\nu}T_{\mu\nu}+\frac{1}{2\Mpl}\pi T\right)\,,
\end{equation}
where $(\mathcal{E}h)_{\mu\nu}=-\frac{1}{2}\bx h_{\mu\nu}+\cdots$ is the Lichnerowicz operator,
and $T$ is the trace of the stress-energy tensor. As we can see,
the helicity-2 and -0 modes decouple in this case\footnote{We point out however that in Ghost-free Massive Gravity the helicity-2 and -0 modes do not fully decouple in the `decoupling limit' unless $c_5=0$. Furthermore that theory also includes other non-conformal couplings to matter which we ignore here. See Refs.~\cite{Wyman:2011mp,Sjors:2011iv} for effects arising from these non-conformal coupling to matter.}.
Here the $\lag_{i}$ are the Galileon interactions in four dimensions
\begin{eqnarray}
\label{eq:l2-lag}
\lag_{2} & \equiv & \left(\pd\pi\right)^{2}\\
\label{eq:l3-lag}
\lag_{3} & \equiv & \left(\pd\pi\right)^{2}[\Pi]\\
\label{eq:l4-lag}
\lag_{4} & \equiv & \left(\pd\pi\right)^{2}\left([\Pi]^{2}-\left[\Pi^{2}\right]\right)\\
\label{eq:l5-lag}
\lag_{5} & \equiv & \left(\pd\pi\right)^{2}\left([\Pi]^{3}-3[\Pi]\left[\Pi^2\right]+2\left[\Pi^3\right]\right)
\end{eqnarray}
where $\Pi_{\mu\nu}\equiv\partial_\mu\partial_\nu \pi$ and $\Pi_{\mu\nu}^n = \Pi_\mu^{\alpha_1}\Pi^{\alpha_2}_{\alpha_1}\cdots \Pi^{\alpha_{n}}_\nu$. Square brackets $[A]$ denote the trace of the tensor $A$ with respect to the Minkowski metric $[A]\equiv\eta^{\mu\nu}A_{\mu\nu}$.

The equations of
motion for $\pi$ and $h_{\mu\nu}$ are then
\begin{eqnarray}
\label{eq:metric-eoms}
(\mathcal{E}h)_{\mu\nu} & = & \frac{1}{\Mpl}T_{\mu\nu} \, ,\\
\label{eq:gal-eoms}
-\frac{3}{4}\sum_{i=2}^{5}\frac{c_{i}}{\Lm_{i}^{3(i-2)}}\frac{\delta\lag_{i}}{\delta\pi} & = & \frac{1}{2\Mpl}T\,.
\end{eqnarray}
The coefficients $c_{i}$ are dimensionless, and so far arbitrary, although $c_2,c_3,c_4$ must be positive for the stability of the theory (see Ref.~\cite{Nicolis:2008in}). Without loss of generality, we can absorb $c_2$ into the definition of $\pi$ and $c_3$ into that of $\Lambda$ (\ie we can set $c_2=1$ and $c_3=1/3$)\footnote{Strictly speaking redefining $c_2$ corresponds to changing the coupling to external matter. Of course by making that coupling small we reduce the amount of radiation into the Galileon, but in their natural realizations the coupling to matter is of order 1 and one is bound to rely on the Vainshtein mechanism to hide this scalar field.}.
The $\Lm_{i}$ are the 
scales associated with each of the Galileon interactions.
These scales are typically assumed to be of the same order $\Lm$ and in theories of ghost-free massive gravity this
scale is related to the mass of the graviton by $\Lm\sim(m^{2}\Mpl)^{1/3}$.
Current bounds on the graviton mass typically require $m$ not to be too much larger than $m\sim H_{0}\sim10^{-33}{\rm eV}$, in which case $\Lm\sim(1000{\rm \ km})^{-1}$.
However, for a generic Galileon theory, one could potentially consider these scales as being different.
The non-renormalization theorem present in Galileon theories allows for such a hierarchy without fine-tuning issues, \cite{Luty:2003vm,Nicolis:2008in}.
The notation used here is similar to that in Ref.~\cite{deRham:2012fw}, after setting $\Lm_{3}=\Lm$,
$c_{2}=1$, $c_{3}=1/3$.
For simplicity we also set  $c_4=1$ in what follows. We leave $c_5$ arbitrary because it can be of either sign.

In the rest of this paper, we will assume that a Vainshtein mechanism does occur and that at short enough distances the interactions \eqref{eq:l3-lag} or \eqref{eq:l4-lag} dominate over the standard kinetic term \eqref{eq:l2-lag}. This depends on the relation between the different coefficients in the theory, which we assume to be a given fact.

If $\Lm_{4}\lesssim\Lm_{3}$ then $\lag_{3}$ never has
any effect because it will only become relevant at energies where
it is already dominated by $\lag_{4}$.
If on the other hand $\Lm_{3}\leq\Lm_{4}$, then the interactions $\lag_{3}$ can dominate for a little while before being taken over by $\lag_{4}$ at short enough distances.
As we will see, it will also be convenient to take $\Lm_{5}\geq(\Lm_{4}/\Lm_{3})^{1/3}\Lm_{4}$.
In the first part of this paper we will have in mind the situation $\Lm_{4}\lesssim\Lm_{3}\sim(1000{\rm km})^{-1}$. Notice that for a spherically symmetric  configuration, the quintic interactions $\Lm_{5}$ vanish, and so these interactions are only relevant at the perturbed level (however as we shall see, even at that level, they simply correspond to a rescaling of some parameters).

Our basic philosophy for computing the power emission is to perform a background+perturbation split in
the Galileon where the background is static and spherically symmetric and
the deviations from spherical symmetry is captured by the perturbations.
In effect this decomposition assumes that the majority of the Vainshtein screening comes from the monopole moment of the binary system.
More precisely
we split the field $\pi$ and the source as
\ba
\pi(\vec{x},t)&=&\pi(r)+\sqrt{2/3}\phi^{(1)}(\vec{x},t)+\cdots\\
T&=&T_0+\delta T\,,
\ea
where $\(\pd\pd \pi(r)\)^3\sim T_0$ and
$\(\pd\pd \pi(r)\)^2 \pd\pd \phi^{(1)}\sim\delta T$, if the interaction $\Lm_{4}$ dominates.

Physically this split is suggested from standard Effective Field Theory considerations, where we expect
that the physics responsible for the radiation should arise at the (energy)
scale $\Op$. Since this is scale is much smaller than the scale associated with
the size of the system $\bar r^{-1}$ (typical distance between the two objects),
where the spherical symmetry is broken, we expect that spherical background should be a good approximation when
computing the power, barring some unusual circumstances.
However we emphasize that we are free to choose any background+perturbation split so long as the resulting perturbative
expansion remains under control.

The rest of this section is organized as follows. In the next two subsections we will solve for the background field $\pi$ and derive the equations of motion for the fluctuations $\phi^{(1)}$. Then we will review how to compute the power using the Feynman propagator constructed from the fluctuations.

\subsection{Static and Spherically Symmetric Background}

Assuming a point source $T_{0\ \ \nu}^{\ \mu}=-\Mtot\delta^{(3)}(\vec{x})\delta_{0}^{\mu}\delta_{\nu}^{0}$, where $\Mtot=M_1+M_2$ is the total mass of the binary system,
the background solution for $\pi$ is spherically symmetric. Using the notation $\vec{\nabla}\pi(r)=\hat{r}E(r)$, the background field equation for $\pi$ takes the simple algebraic form
\begin{equation}
\label{eq:background-eom}
\left(\frac{E}{r}\right)+\frac{2}{3\Lm_{3}^{3}}\left(\frac{E}{r}\right)^{2}+\frac{2}{\Lm_{4}^{6}}\left(\frac{E}{r}\right)^{3}
=\frac{1}{12\pi }\frac{\Mtot}{\Mpl}\frac{1}{ r^{3}}\,.
\end{equation}
The quintic Galileon does not affect the background configuration, \cite{Nicolis:2008in}. This is because the $k$-th Galileon term is a topological
invariant in dimensions smaller than $k-1$, and since the system
is static it is effectively three dimensional.
One therefore has three branches of solution.
We focus here on the `normal' branch, which smoothly connects a free (weakly interacting) field $E \sim 1/r^2$
at spatial infinity $r\rightarrow\infty$ to a strongly interacting field at short distance scales, so as to achieve the Vainshtein mechanism.

The source has two Vainshtein radii, $\rstarThree $ and $\rstarFour$ associated to the two interaction scales $\Lambda_3$ and $\Lambda_4$,
\begin{eqnarray}
\label{eq:def-rstarThree}
\rstarThree  & \equiv & \left(\frac{\Mtot}{16\Mpl}\right)^{1/3}\frac{1}{\Lm_{3}}\\
\label{eq:def-rstarFour}
\rstarFour & \equiv & \left(\frac{\Mtot}{16\Mpl}\right)^{1/3}\left(\frac{\Lm_{3}}{\Lm_{4}}\right)^{3}\frac{1}{\Lm_{4}}\,.
\end{eqnarray}
These two radii define three regimes in space, where $\lag_{2,3,4}$
dominate in turn. More precisely, $E(r)$, which is just the radial
derivative of the background solution, is given by
\begin{eqnarray}
\label{eq:background-soln-big-r}
&& E(r\gg \rstarThree ) =  \frac{\Mtot/\Mpl}{12\pi }\frac{1}{r^{2}}\\
\label{eq:background-soln-mid-r}
&& E(\rstarFour\ll r\ll \rstarThree )  =  \frac{\left(\Mtot/\Mpl\right)^{1/2}}{\sqrt{8\pi r}}\Lambda_3^{3/2}\\
\label{eq:background-soln-small-r}
&& E(r\ll \rstarFour)  =  \frac{(\Mtot/\Mpl)^{1/3}}{(24\pi )^{1/3}}\Lm_{4}^{2}\,.
\end{eqnarray}

\subsection{Equations of motion for perturbations}

The perturbed stress energy tensor $\delta T^{\mu\nu}$ which encodes
the time dependent dynamics, which for slowly moving sources is
\begin{equation}
\label{eq:def-deltaT}
\delta T_{\nu}^{\mu}=-\left[\sum_{i=1,2}M_{i}\delta^{(3)}(\vec{x}-\vec{x}_{i}(t))-\Mtot\delta^{3}(\vec{x})\right]\delta_{0}^{\mu}\delta_{\nu}^{0}\,,
\end{equation}
where $M_{i}$ is the mass of each companion\footnote{See Refs.~\cite{deRham:2012fw,Chu:2012kz} for relativistic corrections to this expression.}.
The quadratic lagrangian for $\phiFirstOrder$ is
\begin{eqnarray}
\label{eq:lag-quad}
&\lag_{\phi} & =  -\frac{1}{2}Z^{\mu\nu}(x)\pd_{\mu}\phiFirstOrder\pd_{\nu}\phiFirstOrder+\frac{\phiFirstOrder}{\sqrt{6}\Mpl}\delta T\\
 & = & \frac{1}{2}Z_{tt}(r)({ \partial_t \phiFirstOrder})^{2}-\frac{1}{2}Z_{rr}(r)\left(\pdr\phiFirstOrder\right)^{2}-\frac{1}{2 r^2}Z_{\Omega\Omega}\left(\pdO \phiFirstOrder\right)^{2}+\frac{\phiFirstOrder}{\sqrt{6}\Mpl}\delta T
\end{eqnarray}
where
\begin{eqnarray}
\label{eq:def-Zrr}
&& Z_{rr}(r)  \equiv 1+\frac{4}{3\Lm_{3}^{3}}\frac{E(r)}{r}+\frac{6}{\Lm_{4}^{6}}\frac{E(r)^{2}}{r^{2}}\\
\label{eq:def-Ztt}
&& Z_{tt}(r)  \equiv  \frac{1}{3r^{2}}\frac{\d}{\d r}\left[r^{3}\left(1+\frac{2}{\Lm_{3}^{3}}\frac{E(r)}{r}+\frac{18}{\Lm_{4}^{6}}\frac{E(r)^{2}}{r^{2}}+\frac{24}{\Lm_{5}^{9}}\frac{E(r)^{3}}{r^{3}}\right)\right]\\
\label{eq:def-ZOO}
&& Z_{\Omega\Omega}(r)  \equiv  \frac{1}{2r}\frac{\d}{\d r}\left[r^{2}\left(1+\frac{4}{3\Lm_{3}^{3}}\frac{E(r)}{r}+\frac{6}{\Lm_{4}^{6}}\frac{E(r)^{2}}{r^{2}}\right)\right]\,.
\end{eqnarray}
For notational consistency we include here $\Lm_{5}$. Recall
however that $\Lm_{5}$ does not contribute at the background
level, and in particular there is no associated Vainshtein radius
$r_{*,5}$. Its only effect is to redress the time derivative pieces
of the action.

As we will see the quadratic action will only be sufficient to compute
the power when there is a large hierarchy between $\Lm_{3}$ and
$\Lm_{4}$ or between $M_{1}$ and $M_{2}$. However for now we
continue on without making any assumptions.
The quadratic action gives rise to the equations of motion
\begin{equation}
\label{eq:linear-eom}
\hat{\bx}\phiFirstOrder=-\frac{1}{\sqrt{6}\Mpl}\delta T\,,
\end{equation}
where $\hat{\bx}$ is the modified d'Alembertian defined as
\ba
\label{dAlembertian}
\hat{\Box}= -\p_t \(Z_{tt}\p_t\)+ \p_r\(Z_{rr}\p_r\)+\frac{1}{r^2 }Z_{\Omega \Omega} \nabla_{\Omega}^2 \, ,
\ea
where $ \nabla_{\Omega}^2$ is the Laplacian on a unit 2-sphere.
It is useful to consider the form of $\hat{\bx}$
in the different regions.

$\bullet$ \textbf{$\lag_{2}$ region: $r\gg \rstarThree $}
\begin{equation}
\label{eq:linear-eom-big-r}
\hat{\bx}\phiFirstOrder=\bx\phiFirstOrder\,,
\end{equation}
\ie far from the source the field is weakly coupled and perturbations
are free as required.

$\bullet$ \textbf{$\lag_{3}$ region: $\rstarFour\ll r\ll \rstarThree $}\\
In this case, so long as $\Lm_{5}\gg(\Lm_{4}/\Lm_{3})^{1/3}\Lm_{4}$
the equations of motion reduce to the normal cubic Galileon equation
of motion
\begin{equation}
\label{eq:linear-eom-mid-r}
\hat{\bx}\phiFirstOrder=\sqrt{\frac{512}{9\pi}}\left(\frac{\rstarThree }{r}\right)^{3/2}\left(-3\pdt^{2}\phi+4\pdr^{2}\phi+\frac{2}{r}\pdr\phi+\frac{1}{r^{2}}\pdO ^{2}\phi\right)\,.
\end{equation}
The case of small $\Lm_{5}$ is considered in Appendix~\ref{sec:App2}.

$\bullet$ \textbf{$\lag_{4}$ region:} $r\ll \rstarFour$
\begin{equation}
\label{eq:linear-eom-small-r}
\hat{\bx}\phiFirstOrder=\frac{128\times3^{1/3}}{\pi^{2/3}}\left(\frac{\Lm_{4}}{\Lm_{3}}\right)^{6}\left(\frac{\rstarFour}{r}\right)^{2}\left[-\frac{1}{c_r^2}\pdt^{2}\phi+\pdr^{2}\phi+\frac{k_{\Omega}}{\rstarFour^{2}}\pdO ^{2}\phi\right]\,,
\end{equation}
where the speed of sound of the radial fluctuations $c_r$ is given by
\begin{equation}
\label{eq:def-cr}
c_r=\left(1-c_5\frac{4}{9}\frac{\Lambda_4^{12}}{\Lambda_3^3 \Lambda_5^9}\right)^{-1/2}\, ,
\end{equation}
and the coefficient $k_\Omega$ is given by
\begin{equation}
\label{eq:def-kO}
k_{\Omega} = \frac{\pi^{2/3}}{1728\times 3^{1/3}}\left(1-\frac{27}{2}\left(\frac{\Lambda_3}{\Lambda_4}\right)^6\right)\, .
\end{equation}
Note that in the $\lag_{4}$ region the fluctuations effectively
see a one dimensional metric $\d s^2=-Z_{\mu\nu}\rmd x^{\mu}\rmd x^{\nu}\propto-\rmd t^{2}+\rmd r^{2}+\rstar ^{2}\rmd \Omega^{2}$,
where crucially the angular part of the metric is multiplied by the
constant $\rstar ^{2}$ instead of the normal factor of $r^{2}$.

Note that the second term is order 1 so long as $\Lm_{5}\geq(\Lm_{4}/\Lm_{3})^{1/3}\Lm_{4}$.
This is the same condition that we found above for the $\lag_{5}$
contribution to be negligible in the $\lag_{3}$ region. The
effect of $\lag_{5}$ here is to decrease the sound speed,
$c_{r}\sim (\Lm_{5}^{9}\Lm_{3}^{3}/\Lm_{4}^{12})$.
This case is considered in Appendix~\ref{sec:App2}.

The stability of these theories was studied  in Ref.~\cite{Nicolis:2008in},
so as long as we take our coefficients to satisfy the conditions of Ref.~\cite{Nicolis:2008in} perturbations
are guaranteed to be stable about the spherically symmetric configuration.

\subsection{Computing the power using the effective action}

Following \cite{Goldberger:2004jt} we compute the power in the binary pulsar system
by looking at the imaginary part of the effective action\footnote{This method differs slightly from that followed in \cite{Chu:2012kz}, but both strategies are valid and are ultimately equivalent.}
obtained by integrating out the fluctuations $\phi$.
We start with the quadratic action for the perturbations
\begin{equation}
\label{eq:quad-action}
S[\phi,x_{i}^{\mu}]=\intdx \sqrt{-g}\left(-\frac{1}{2}Z^{\mu\nu}\pd_{\mu}\phi\pd_{\nu}\phi+\frac{1}{\sqrt{6}\Mpl}\phi\delta T\right)\,,
\end{equation}
where the $x_{i}^{\mu}$ are the trajectories of the two objects. We then
integrate out the field $\phi$ leaving us with an effective action
\begin{eqnarray}
\label{eq:eff-action}
S_{{\rm eff}} & = & \intdx \ \mathcal{L_{M}}+\frac{i}{12\Mpl^{2}}\int \d^{4}x \d^{4}x'\delta T(x)G_{F}(x,x')\delta T(x')\\
 & + & \text{usual helicity-2 contributions from GR}\,.\nn
\end{eqnarray}
Here we have used the fact that the field $\phi$ can be expressed
in terms of the Feynman propagator
\begin{equation}
\label{eq:prop-soln}
\phi(x)=\frac{i}{\sqrt{6}\Mpl}\intdx 'G_{F}(x,x')\delta T(x')\,,
\end{equation}
where we have defined the Feynman propagator
\begin{equation}
\label{eq:def-feynman}
\hat{\bx}G_{F}=i\delta^{4}(x-x')\,,
\end{equation}
and where the modified d'Alembertian operator $\hat{\bx}$ is defined
in \eqref{dAlembertian}.

As usual the Feynman propagator can be expressed in terms of the Wightman
functions
\begin{equation}
\label{eq:feynman-in-wightman}
G_{F}(x,x')=\theta(t-t')W^{+}(x,x')+\theta(t'-t)W^{-}(x,x')\,,
\end{equation}
where
\begin{equation}
\label{eq:def-wightman}
W^{+}(x,x')=\sum_{\ell m}\int_{0}^{\infty}\rmd \omega\ u_{\ell\omega}(r)u_{\ell\omega}^{*}(r')Y_{\ell m}(\theta,\varphi)Y_{\ell m}^{*}(\theta',\varphi')e^{-i\omega(t-t')}\,,
\end{equation}
and the mode functions expanded in spherical harmonics $u_{\ell m \omega}(r,\Omega, t)=u_{\ell\omega}(r)Y_{\ell m}(\Omega)e^{-i\omega t}$ are the eigenfunctions of the mode equation $\hat \Box u_{\ell m \omega}=0$ and form a complete set. \\ %
The time averaged power is
\begin{equation}
\label{eq:def-power}
P=-\Big<\frac{dE}{dt}\Big>=\int_{0}^{\infty}\rmd \omega\ \omega f(\omega)\,,
\end{equation}
where $f(\omega)$is determined from the imaginary part of the effective
action
\begin{equation}
\label{eq:def-f}
\frac{2}{\Tp}{\rm Im} S_{{\rm eff}}=\int_{0}^{\infty}\rmd \omega f(\omega)\,.
\end{equation}
We define the moments
\begin{equation}
\label{eq:def-Mom}
\Mom _{\ell mn}=\frac{1}{\Tp }\int_{0}^{\Tp }\rmd t\int\rmd ^{3}x\ u_{\ell n}(r)Y_{\ell m}(\theta,\varphi)e^{-in\Op t}\delta T(\vec{x},t)\,,
\end{equation}
where we use the notation $u_{\ell n}\equiv u_{\ell,\, \omega=n \Omega}$.
Taking the Fourier transform
\begin{equation}
\label{eq:ft-of-Mom}
\Mom _{\ell m}=\sum_{n=-\infty}^{\infty}\Mred _{\ell mn}e^{in\Op t}\,,
\end{equation}
we have
\begin{equation}
\label{eq:f-solved}
f(\omega)=\frac{1}{3\Mpl^{2}\Tp }\sum_{\ell m}\int_{0}^{\Tp }\rmd t\int_{-\infty}^{t'}\rmd t'{\rm Re}\left(e^{i\omega(t-t')}\Mred _{\ell m}(t)\Mred _{\ell m}^{*}(t')\right)\,,
\end{equation}
where we have used the facts that the imaginary part of $u_{\ell m}$ integrates to zero and that $\sum_{m=-\ell}^{\ell}Y_{\ell m}(\theta,\varphi)Y_{\ell m}(\theta',\varphi')$
is real. Then the period-averaged power emission is
\begin{equation}
\label{eq:power-solved}
\la P\ra=\frac{\pi}{3\Mpl^{2}}\sum_{n=0}^{\infty}\sum_{\ell m}n\Op |\Mred _{\ell mn}|^{2}\,.
\end{equation}
Notice that since we only integrate over positive frequencies in \eqref{eq:def-power}, we only need to sum over positive harmonics $n\ge 0$.

\section{Power Emitted in the Quartic Galileon}
\label{sec:Power_1}

First we consider the case that there is a single strong coupling
scale, $\Lm_{3}=\Lm_{4}=\Lm_{5}\equiv\Lm$, so $\rstarThree =\rstarFour\equiv \rstar $.
We compute the power emitted in the Galileon using the formalism
developed in the last section. This requires us to derive the properly normalized mode functions
$u_{\ell n}$.

\subsection{Mode functions}

The field fluctuations $\phi$ can be expanded in terms of the mode functions
\begin{equation}
\label{eq:phi-expansion}
\phi(x,t)=\sum_{n=-\infty}^{\infty}\sum_{\ell m}a_{\ell mn}u_{\ell n}(r)e^{-in\Op t}Y_{\ell m}(\Omega)\,,
\end{equation}
where we have used the fact that for periodic systems we need only
sum over a discrete set of harmonics $n$ instead of integrating over
a continuum of frequencies.

The radial mode functions $u_{\ell n}$ are solutions to the homogeneous
equation
\ba
\hat{\bx}u_{\ell n}e^{-in\Op t}Y_{\ell m}(\Omega)=0\,,
\ea
subject to the normalization defined by equations \eqref{eq:def-feynman} and \eqref{eq:def-wightman}, which is valid as long as the field reaches the oscillating WKB regime within the strong coupling region (\ie as long as $R_\star> \Omega^{-1}$).
In practice however we compute this normalization by matching with the free Minkowski spacetime normalization at spatial infinity,
$r\gg \rstar$
\begin{equation}
\label{eq:norm-big-r}
\lim_{r\rightarrow\infty}u_{\ell n}(r)=\frac{1}{\sqrt{\pi\omega}}\frac{\cos(n\Op r+P)}{r}\,,
\end{equation}
where the phase $P$ is irrelevant.

\subsubsection{Strong Coupling Regime}

In the strong coupling region where $\lag_{4}$ dominates, $r\ll \rstar $, the mode functions
which satisfy the correct boundary conditions at the origin are (see Appendix~\ref{sec:App1} for
a detailed discussion of the choice of boundary condition)
\begin{eqnarray}
\label{eq:monopole-norm}
u_{0n}(r) & = & \bar{u}_{0n}\cos\omega_{0n}r\ \ \ \ \ \ \ (\ell=0)\\
\label{eq:multipole-norm}
u_{\ell n}(r) & = & \bar{u}_{\ell n}\sin\omega_{\ell n}r\ \ \ \ \ \ \ (\ell>0)\,,
\end{eqnarray}
where
\begin{equation}
\label{def-omega-l}
\omega_{\ell n}^{2}\equiv \frac{1}{c_r^2}(n\Op )^{2}-k_{\Omega}\frac{\ell(\ell+1)}{\rstar^{2}}\,,
\end{equation}
and where the normalization $\bar{u}$ is fixed for each mode by matching onto the
correctly normalized mode at infinity according to \eqref{eq:norm-big-r}.

Matching the WKB solution with the strong coupling solution at $r=\rstar $
we find
\begin{equation}
\label{eq:ubar-solved}
\bar{u}_{\ell n}=\begin{cases}
\frac{1}{\sqrt{\pi n\Op }\rstar }\,, & \ell\ll n\lcrit \\
\frac{e^{-\ell^{2}}}{\sqrt{\pi n\Op }\rstar }\,, & \ell\gg n\lcrit
\end{cases}\,,
\end{equation}
where $\lcrit $ is defined by
\begin{equation}
\label{eq:def-lcrit}
\lcrit \equiv \frac{1}{c_r \sqrt{k_\Omega}}\, \Op \rstar\,.
\end{equation}
Modes with imaginary $\omega_{\ell n}$ are exponentially suppressed,
so when computing the power we only need to sum over modes with real $\omega_{\ell n}$.
With this in mind we use the following approximation in what follows,
\ba
\bar{u}_{\ell n}=\theta(n\lcrit -\ell)\frac{1}{\sqrt{\pi n\Op }}\,,
\ea
where we define the step function as $\theta(x)=0$ for $x<0$ and 1 otherwise.

\subsubsection{General Form of the Power}

As usual, by conservation of energy the monopole does not radiate in the non-relativistic limit (\ie at leading order in the velocity expansion. See Ref.~\cite{deRham:2012fw,Chu:2012kz} for the behaviour at next order). Similarly, by momentum conservation the dipole does not radiate at leading order in the velocity expansion\footnote{
Normally one associates the dipole moment with the moment linear in the velocity $v$ and by momentum conservation this moment does not radiate. In our case whilst it is true that the dipole moment defined as $\ell=1$ does not radiate,
the higher order multipoles also have only one power of $v$, and they do radiate.
The spherical harmonics with $\ell>1$ allow for time dependence.}.
The power for the higher order multipoles is given by
\begin{eqnarray}
\label{eq:power-multipole}
\la P\ra & = & \frac{\pi}{3\Mpl^{2}}\sum_{n=0}^{\infty}\sum_{\ell=1}^{\infty}\sum_{m=-\ell}^{\ell}n\Op \left(\frac{\theta(\ell-n\lcrit )}{\sqrt{\pi n\Op }\rstar }\right)^{2}\\
&&\times\left|\frac{1}{\Tp }\int_{0}^{\Tp }\rmd t\ \int\rmd ^{3}x\ e^{-in\Op t}\sin(n\Op r)Y_{\ell m}(\theta,\phi)
\delta T(\vec{x},t)\right|^{2}\nn\,,
\end{eqnarray}
where the expression for the perturbed source $\delta T(\vec{x},t)$ is given in \eqref{eq:def-deltaT} and is proportional to $\delta^{(3)}(\vec{r}-\vec{r}_{i}(t))$, and the two objects follow the standard Keplerian orbits $\vec{r}_{i}(t)$,
given in spherical coordinates by
\begin{eqnarray}
\label{eq:def-ri}
r_{1,2}(t) & = & \frac{M_{2,1}}{\Mtot}\frac{\bar{r(}1-\ep ^{2})}{1+\ep \cos\Op t}\\
\label{eq:def-thetai}
{\rm with }\hspace{10pt} \theta_{1,2}(t) & = & \frac{\pi}{2}\hspace{10pt}{\rm and }\hspace{10pt}
\varphi_{1,2}(t) =  \Op t+\delta_{i,2}\pi\,,
\end{eqnarray}
where $\ep $ is the eccentricity and $\rbar $ is the semi-major
axis of the orbit. For simplicity (and without loss of generality), we choose the plane of the orbits to be localized in the plane $\theta\equiv \pi/2$. The remaining angle $\varphi$ is then determined knowing that the two objects orbit with frequency $\Op $ and are always diametrically opposed on their orbital path.

Defining the reduced mass
\begin{equation}
\label{eq:def-Mred}
\Mred \equiv\frac{M_{1}M_{2}}{\Mtot}\,,
\end{equation}
then to leading order in $\Op \rbar $ the power emitted is
\begin{eqnarray}
\label{eq:power-order-v1}
\la P\ra&=&\frac{\Mred ^{2}\Op ^{2}\rbar ^{2}}{3\Mpl^{2}\rstar ^{2}}\sum_{n=0}^{\infty}\sum_{\ell m}\theta(\ell-n\lcrit )n^{2}Y_{\ell m}\left(\frac{\pi}{2},0\right)^{2}\\
&&\times\left|\frac{1+(-1)^{m}}{\Tp }\int_{0}^{\Tp }\rmd t\ e^{-i(n-m)\Op t}\frac{1-\ep ^{2}}{1+\ep \cos\Op t}\right|^{2}\,.\nn
\end{eqnarray}
Note that in Minkowski spacetime radiation problems, the mode functions are $j_{\ell}(\omega r)\approx(\omega r)^{\ell}$
at small distances, and so higher order multipoles are suppressed by
more powers of $\Op \rbar \sim v$. Here however there is no
additional velocity suppression for higher order multipoles (for as $\ell < n \lcrit$).

The integral can be evaluated as
\begin{equation}
\label{eq:eccentricity-integral}
\frac{1}{\Tp }\int_{0}^{\Tp }\rmd t\frac{e^{-i(n-m)\Op t}}{1+\ep \cos\Op t}=(-1)^{n-m}\frac{2\pi}{\sqrt{1-\ep ^{2}}}\left(\frac{\ep }{1+\sqrt{1-\ep ^{2}}}\right)^{n-m}\,,
\end{equation}
for $n-m\ge 0$,
so the power is
\begin{eqnarray}
\label{eq:power-after-eccentricty-integral}
\la P\ra&=&\frac{8\pi^{2}}{3}\(\frac{\Mred }{\Mpl}\)^2\left(\frac{\rbar }{\rstar }\right)^{2}\Op ^{2}\, S_{\ep}\,,
\end{eqnarray}
with
\ba
\label{sum}
S_{\ep }=\sum_{n=0}^{\infty}\sum_{\ell=0}^{n\lcrit }\sum_{m=-\ell}^{\ell}n^{2}Y_{\ell m}\left(\frac{\pi}{2},0\right)^{2}\left(1-\ep ^{2}\right)\left(\frac{\ep }{1+\sqrt{1-\ep ^{2}}}\right)^{2(n-m)}\cos^{2}\left(\frac{m\pi}{2}\right)\nn \,.
\ea
Note that the $\ell=1$ mode does not radiate (since $Y_{1,0}(\pi/2,0)=0$), as expected from momentum conservation.
So the first multipole that has nonzero radiation is the quadrupole $\ell=2$.

\subsubsection{Quadrupole}

Let us now compare the power emitted in the quadrupole to the power
from GR and from the cubic Galileon (in the case where the quartic and quintic interactions are absent, but keeping the same strong coupling scale $\Lambda$).
The power emitted by the quadrupole in the $n=1$ harmonic is
\begin{equation}
\label{eq:power-l4-quad}
\la P\ra_{{\rm Full\ Galileon}}^{(\ell=2)}\sim\left(\frac{\Mred }{\Mpl}\right)^{2}\frac{(\Op \rbar )^{2}}{(\Op \rstar )^{2}}\, \Op ^{2}\,.
\end{equation}
Comparing this result with that of the cubic Galileon presented in Ref.~\cite{deRham:2012fw},
\begin{equation}
\label{eq:power-l3-quad}
\la P\ra_{{\rm Cubic\ Galileon}}^{(\ell=2)}\sim\(\frac{\mathcal{M}}{\Mpl}\)^{2}\frac{(\Op \rbar )^{3}}{(\Op \rstar )^{3/2}}\, \Op ^{2}\,.
\end{equation}
We see that the relevant Vainshtein screening is $(\Op \rstar )^{-2}$
compared to the Vainshtein screening appropriate for the force between
the pulsars $(\rbar /\rstar )^{2}$. This is exactly analogous to what
happens in the cubic Galileon, where the Vainshtein screening is less
effective than in the static case. We also see that the full  Galileon
is enhanced by a factor $(\Op \rbar )^{-1}\sim v^{-1}$ relative to the cubic Galileon
because the quadrupole is sourced by the monopole
moment.

So far, we have only considered the contribution from the first harmonic.
The Galileon radiation includes radiation from all harmonics, each
of which contributes equally to the moment.

\subsubsection{Summing over all multipoles and harmonics}

If one were allowed to sum over all harmonics (till $n\to \infty$),
the power emitted would formally diverge. In order to gain a better understand of this divergence we truncate the sum over the harmonics at the cutoff $\nmax$ in \eqref{sum}, and denote as $S_{\ep}(\nmax)$ this regularized sum.
Furthermore, for simplicity we focus on the case with no eccentricity, \ie $\ep =0$, so that the regularized sum simplifies to
\ba
\label{eq:regularized-sum}
S_{0}(\nmax)=\sum_{n=0}^{\nmax}\sum_{\ell=n}^{n\lcrit }\sum_{m=-\ell}^{\ell}n^{2}Y_{\ell m}\left(\frac{\pi}{2},0\right)^{2}\cos^{2}\left(\frac{n\pi}{2}\right) \,.
\ea
Since $\Op r_\star \sim 10^6 \gg 1$ for realistic pulsars, most of the terms in the sum have $\ell \gg n$. We can then use the approximation valid for $\ell \gg n$,
\begin{equation}
\label{eq:Ylm_approx}
Y_{\ell n}\(\frac{\pi}{2},0\)\approx\frac{1}{\pi}\cos\left((\ell+n)\frac{\pi}{2}\right)\,.
\end{equation}
The details of the calculation can be found in Appendix~\ref{sec:App3}, and the final result is
\begin{equation}
\label{eq:power-answer}
\la P\ra\approx\frac{1}{12}\(\frac{\Mred }{\Mpl}\)^2\, \left(\frac{\rbar }{\rstar}\right)^{2}\lcrit \nmax^{4}\, \Op ^{2}\,,
\end{equation}
the power depends quartically on the cutoff $\nmax$.

To get a sense of this result we apply it to a pulsar system consisting of two solar mass objects orbiting with a period of $2\pi/\Op=8$ hours with a semi-major axis $\bar{r}=10^9$m and with no eccentricity. This choice of parameters is close to those of the Hulse-Taylor pulsar \cite{Weisberg:2004hi}.

The GR result for this system is given by the Peters-Mathews formula (assuming zero eccentricity)
\be
P_{\rm Peters-Mathews}=\frac{2^{5}}{5}\frac{M_{1}^2 M_{2}^2 \Mtot}{\bar{r}^5}\,.
\ee
Comparing this to the naive Galileon result \eqref{eq:power-l3-quad} we find
\be
\frac{P_{\rm Full\ Galileon}}{P_{\rm Peters-Mathews}}\approx  6\times 10^{-4}\ \nmax^{4}\,.
\ee
Even for $\nmax\sim 1$ this is a large amount of power compared to the cubic Galileon case in Eq.~\ref{eq:power-l3-quad}.
However if we trust this calculation then there is no natural cutoff in $n$ until $n\sim (\Op \bar{r})^{-1}$ when the assumption we made that $\omega \bar{r}\ll 1$ breaks down. If we take this cutoff and
use Hulse Taylor parameters we find that $\frac{\la P_{\rm Full\ Gal}\ra}{\la P\ra_{\rm GR}}\sim10^{9}$. In what follows we will interpret this result as a breakdown in perturbation theory.

\subsection{Validity of Perturbation Theory}

The divergent power suggests that our calculation was too naive. Since we have been using linearized perturbation
theory, the natural thing to check is whether the fluctuations themselves
become nonlinear. Indeed, we might expect perturbation theory to break
down on physical grounds. We have used perturbation theory around
a spherically symmetric source, but have found important contributions
from arbitrarily high multipoles $\ell$. Since higher values of $\ell$ are
more sensitive to what happens over small angles, we expect that our choice of background should become worse for large $\ell$.

In this section we check the validity of perturbation theory around the spherically symmetric background, by explicitly constructing the first and second order perturbations. Since we want to compare the physical values of the fields,
we use here the retarded propagator.

Based on the discussion above we cutoff the sum over $\ell$ in the propagator. If one can trust the perturbation series
at all, it can only be trusted at low $\ell$. Physically this is because there is some uncertainty
associated with the angular position of the two objects. Trusting perturbation
theory to arbitrarily high $\ell$ and performing the sum over all $\ell$
would imply that the positions are known with arbitrary accuracy. High
$\ell$ modes are also more sensitive to the non spherical nature
of the source, and at high enough $\ell$ we do expect the assumption of a spherical
background to break down. Thus we introduce a cutoff $L$ on the sum
over $\ell$. We take the cutoff $L<\lcrit $, since $\lcrit $ is itself very large for realistic pulsar systems.

The analysis is performed in the WKB regime where derivatives
acting on $\phiFirstOrder$ can be expanded in powers of $\Op ^{-1}$.
As we shall see below, the field diverges at certain isolated points and we compare the field values at these points because they will give the largest values of $\phiSecondOrder/\phiFirstOrder$. Finding this ratio to be bigger than one at a single point is sufficient to show that perturbation theory is breaking down.


The exact equations of motion for the Galileon $\pi(x)$ is
\ba
\label{eq:eom-pi}
&&
\bx\pi+\frac{1}{\Lm^3}\( \(\bx\pi\)^2-\(\pd_\mu \pd_\nu \pi\)^2 \)\nn \\
&& +\frac{1}{\Lm^{6}}\left[\left(\bx\pi\right)^{3}-3\bx\pi\left(\pd_{\mu}\pd_{\nu}\pi\right)^{2}
+2\left(\pd_{\mu}\pd_{\nu}\pi\right)^{3}\right]=-\frac{T}{3\Mpl}\,.
\ea
Performing a background/perturbation split for the source, $T=T_{0}+\delta T$,
we will be interested in the second order  perturbations, for the field $\pi=\pi(r)+\sqrt{2/3}\phiFirstOrder+2/3\phiSecondOrder+\cdots$,
with $T_{0}\sim\pi(r)$ and $\delta T\sim\phiFirstOrder$, as already computed in the previous section, (in  particular in the strong coupling region the background configuration for $\pi(r)$ is given in \eqref{eq:background-soln-small-r}).
The second order fluctuation $\phiSecondOrder$ is sourced by nonlinearities in $\phiFirstOrder$.
To check the validity
of perturbation theory, we will compare the magnitude of $\phiSecondOrder$
to the magnitude of $\phiFirstOrder$.
The perturbative expansion is under control only if $\phiFirstOrder\gg\phiSecondOrder$ for all $r$.

Unlike in the previous section where we were interested in the power emitted, we here focus on the physical values of the fields,
and thus construct the retarded propagator,
\begin{eqnarray}
\label{eq:retarded-prop}
G_{R}(x,x') & = & -i\theta(t-t')\la0\big|[\phi(x),\phi(x')]\big|0\ra\\
 & = & -\theta(t-t')\int d\omega\sin\omega(t-t')\\
 &&\times\sum_{\ell=0}^{L}\sum_{m=-\ell}^{\ell}u_{\ell\omega}(r)u_{\ell\omega}(r')Y_{\ell m}(\theta,\varphi)Y_{\ell m}^{*}(\theta',\varphi')\nn\,.
\end{eqnarray}
In the WKB regime, the first order field fluctuation $\phiFirstOrder$ is then given by
\ba
\label{eq:phi1-solved}
\phiFirstOrder(x)&=&-\intdx G_R(x,x')\frac{\delta T(x')}{\sqrt{6}\Mpl}\\
&=&\frac{2\pi\Mred \rbar }{\sqrt{6}\Mpl \rstar ^{2}}\sum_{\ell=1}^{L}\sum_{m=0}^{\ell}\theta(\ell-m\lcrit )\\
&&\times Y_{\ell m}\(\theta,0\)Y_{\ell m}\(\frac{\pi}{2},0\)\, \sin m\Op r\sin(m\Op t-m\varphi)\cos^{2}\(\frac{m\pi}{2}\)\nn\,,
\ea
where the sum over the multipoles has been truncated at the cutoff $L$,
see Appendix~\ref{sec:App4} for details of the above calculation.

The fluctuation $\phiFirstOrder$ reaches its maximal value on a radial light cone when $r=t=(2k+1)\pi/2\Op$ with $k\in \mathbb{Z}$, with $\theta=\pi/2$ and $\varphi=0$, (note that $m$ is then forced to be even).
Calling this set of parameters $x_{\rm max}$ (note that $x_{\rm max}$ is not a unique set of parameters, it is just any choice on the radial light cone that satisfies these conditions). If one were to push $L\to \infty$, the sum for $\phiFirstOrder$ would diverge exactly at $x_{\rm max}$. However, in terms of the finite cutoff $L$, the value of the field at $x_{\rm max}$ is given by
\begin{equation}
\label{eq:phi1-at-xmax-with-approx-ylm}
\phiFirstOrder(x_{\rm max},L)\approx  \frac{1}{4.5}\frac{2\pi\Mred \rbar }{\sqrt{6}\Mpl \rstar ^{2}}\frac{L^{2}}{\pi^{2}}=\frac{4}{9\pi\sqrt{6} }\frac{\Mred}{\mpl}\frac{\rbar}{\rstar ^{2}}L^{2}\,,
\end{equation}
where the factor of $1/4.5$ is an approximation\footnote{Naively one would guess
this factor is 1/4 because we are taking roughly half of the $m$
terms and half of the $\ell$ terms. Numerically it seems that $4.5$
for this factor is a better fit.}.

The crucial observation is that $\phiFirstOrder$ does not fall off with
increasing $r$. This is ultimately tied into the fact that the effective
metric for the fluctuations is one dimensional for $r\ll\rstar $. The
result of this is that $\phiSecondOrder$ does not see a compact source
of size $\rbar $ or even $\Op ^{-1}$ as might be expected,
but rather one that extends out to $\rstar $. This pumps a large amount
of energy into the second order perturbation, and this is what
ultimately makes perturbation theory break down.

The perturbed equation for $\phiSecondOrder$ is
\begin{eqnarray}
\label{eq:eom-phi2}
\hat{\bx}\phiSecondOrder & = & \frac{3}{\Lm^{6}}\Big[\bx\pi_{0}\left(\bx\phiFirstOrder\right)^{2}-2\bx\pi_{0}\left(\pd_{\mu}\pd_{\nu}\phiFirstOrder\right)^{2}\\
&&-\pd_{\mu}\pd_{\nu}\pi_{0}\pd^{\mu}\pd^{\nu}\phiFirstOrder\bx\phiFirstOrder+2\pd_{\mu}\pd^{\nu}\pi_{0}\pd_{\nu}\pd^{\lambda}\phiFirstOrder\pd_{\lambda}\pd^{\mu}\phiFirstOrder\Big]\nn\\
&&+ \frac{1}{\Lm^{3}}\left[\left(\bx\phiFirstOrder\right)^{2}-\left(\pd_{\mu}\pd_{\nu}\phiFirstOrder\right)^{2}\right]\,,\nn
\end{eqnarray}
where $\hat{\bx}$ is given by \eqref{dAlembertian}.

The source for the second order fluctuations is in principle complicated, but we simplify its expression by working in the WKB regime $r>\Op ^{-1}$, and focusing on the leading terms in $\Op r$.  The first order field fluctuation then takes the form
\ba
\phiFirstOrder(\vec{x},t)\sim A(r)B(\theta,\varphi)\cos(n\Op t+P_{t})\cos(n\Op r+P_{r})\,,
\ea
where $A(r)$ is a slowly-varying function of $r$ (which varies over distances much
bigger than $\Op ^{-1}$) and $P_{t,r}$ are irrelevant phases.
For $\ell<\lcrit =\Op \rstar $, we have $\pdr^2\phiFirstOrder\sim\pdt^2\phiFirstOrder\sim\sum_{n}(n^{2}\Op ^{2}+\frac{n\Op }{r}+\cdots)\phiFirstOrder\gg\frac{1}{\rstar ^{2}}\pdO \phiFirstOrder$,
so we can ignore the angular derivatives
and focus only on the leading order contribution from the radial and
time derivatives.
Recalling that $\pd^{2}\pi_{0}\sim M^{1/3}\Lm^{2}/r$, the different contributions sourcing $\phiSecondOrder$ in \eqref{eq:eom-phi2} are then of the form
\ba
\frac{1}{\Lambda^6}(\p^2 \pi_0)(\p^2 \phiFirstOrder)^2&\sim&\frac{1}{\Lm^{6}}\frac{M^{1/3}\Lm^{2}}{r}\left[\sum_{n}\left((n\Op )^{2}+\frac{n\Op }{r}+\cdots\right)\phiFirstOrder\right]^{2}\label{eq:lag4-terms} \\
&\sim&\frac{1}{\Lm^{3}}\frac{\rstar }{r}\left(\sum_{nn'}n^{2}n'^{2}\Op ^{4}+\frac{n^2 n'\Op^3}{r}+\cdots\right)(\phiFirstOrder)^{2}\,,\nn \\
\label{eq:lag3-terms}
{\rm and}\ \ \frac{1}{\Lambda^3}(\p^2 \phiFirstOrder)^2&\sim&\frac{1}{\Lm^{3}}\left(\sum_{nn'}n^{2}n'^{2}\Op ^{4}+\cdots\right)\left(\phiFirstOrder\right)^{2}\,.
\ea
The $\frac{1}{\Lm^{3}}\frac{\rstar }{r}\Op^{4}\phi^{2}$ contribution
in \eqref{eq:lag4-terms} should clearly be the dominant one,
however a straightforward calculation show that it actually vanishes exactly, and the contribution from \eqref{eq:lag4-terms} is thus of the same order as that of \eqref{eq:lag3-terms}.

The next order corrections are the $\frac{1}{\Lm^{3}}\frac{\rstar \omega^{3}}{r^{2}}\phi^{2}=\frac{\omega^{4}}{\Lm^{3}}\frac{\rstar }{\omega r^{2}}\phi^{2}$ from the cross term in \ref{eq:lag4-terms} (that arises from $\lag_{4}$) and the $\frac{\omega^{4}}{\Lm^{3}}\phi^{2}$ contribution coming from \ref{eq:lag3-terms} (that arises from $\lag_{3}$).
We see that at small $r$, the $\lag_{4}$
contribution is dominant (as expected), but becomes subdominant when
$r>(\rstar/\omega)^{1/2}=\frac{1}{\sqrt{n}}(\Op \rstar )^{1/2}\Op^{-1}=\frac{1}{\sqrt{n}}(\Op \rstar )^{-1/2}\rstar $,
\ie still within the strong coupling regime but already in the WKB region. This is true for all $n$. Since we are interested in evaluating the source at $r=\rstar $, the contribution from $\lag_{3}$ is the most significant.
This is explained in more depth in Appendix~\ref{sec:App4}.

We can solve the equation for $\phiSecondOrder$ by using a WKB-like ansatz. In the limit $L\to \infty$,  we would find again a diverging expression on the radial light cone, however keeping a fixed cutoff $L$ we find
\begin{equation}
\label{eq:phi-2-at-xmax-after-sum}
\phiSecondOrder(x_{\rm max},L)=\left(\frac{r}{\rstar }\right)^{2}\frac{2}{\Lm^{3}}\left(\frac{2\pi\Mred \rbar }{\sqrt{6}\Mpl \rstar ^{2}}\right)^{2}\Op ^{2}\frac{L^{6}}{121\pi^{4}}\,.
\end{equation}
We can then explicitly compare $\phiFirstOrder$ and $\phiSecondOrder$
\begin{eqnarray}
\label{eq:phi12-ratio-solved}
\frac{\phiSecondOrder(\rstar ;x_{\rm max},L)}{\phiFirstOrder(\rstar ;x_{\rm max},L)} & = & \frac{3\sqrt{6\pi}}{121}\frac{\Mred }{\Mpl}v\frac{1}{(\Lm \rstar )^{2}}\frac{\Op }{\Lm}L^{4}\\
&=& \frac{3}{968\pi}\sqrt{\frac{3}{2}}\frac{\Mred }{\Mtot }v\Op \rstar L^{4} \label{eq:estimatePower}\\
 & \approx & 0.1\times v\Op \rstar L^{4}\,,
\end{eqnarray}
with the orbital velocity $v$ given by $v=\Op \bar r$, where we assumed $\Mred \approx \Mtot$, in the last expression, \ie two bodies of comparable masses.

We find that in principle there are systems where perturbation
is valid over some range of $L$ so long as the system is sufficiently light or slow (small $v\Op \rstar$), \ie as long as
\ba
\label{Cutoff}
L\lesssim \(\frac{\Mred }{\Mtot }v\Op \rstar \)^{-1/4}\,.
\ea
However for realistic binary pulsar systems, $v\Op \rstar \sim10^{3}$, so perturbation theory breaks down for all $\ell$.

\subsection{Effect of a Multipole Cutoff on the Power Emitted}

As mentioned previously, one can only trust perturbation theory as long as the second order field fluctuations are small relative to the first order one (in principle this is not a sufficient condition, but it already gives a good handle on the behaviour of perturbation theory in a given setup). We thus consider a cutoff $L$ which small enough and replace $S_{0}(\nmax)$ in \eqref{eq:regularized-sum} with $S_{0}(L)$, that is to say we replace the cutoff in $n$ with a cutoff in $\ell$ with $L<\lcrit $.
Recomputing the expression for the power with this cutoff, we get
\begin{eqnarray}
\label{eq:sum-with-l-cutoff}
S_{0}(L)  =  \frac{1}{\pi^{2}}\sum_{n=0}^{\infty}\sum_{\ell=1}^{L}\sum_{m=-\ell}^{\ell}n^{2}Y_{\ell m}\left(\frac{\pi}{2},0\right)^{2}\delta_{n,m}
 \approx \frac{L^{4}}{18\pi^{2}}\,,
\end{eqnarray}
where the approximation is obtained numerically\footnote{If we had used
the approximation $\ell\gg m$ we could have done the sum exactly
and would have found $L^{2}/48\pi^{2}$.}. The factor of $m^{2}$ weights
the high $m$ modes (where $m\approx\ell$ more heavily and so this
approximation breaks down).
In terms of the power emitted this implies
\begin{eqnarray}
\label{eq:power-with-l-cutoff}
\la P\ra & = & \frac{8\pi^{2}}{3}\frac{\Mred ^{2}}{\Mpl^{2}}\left(\frac{\rbar }{\rstar }\right)^{2}\Op ^{2}S_{0}(L)\\
 & = & \frac{4}{27}\frac{\Mred ^{2}}{\Mpl^{2}}\frac{v^{2}}{(\Op \rstar )^{2}}\Op ^{2}L^{4}\\
 &=&  \frac{4}{27}\frac{121}{3\sqrt{6\pi}}\frac{\Mred ^{2}}{\Mpl^{2}}\frac{v}{(\Op \rstar )^{3}}\Op ^{2}\,.
\end{eqnarray}
Note that this result is more Vainshtein suppressed than any one mode ($P_{\rm tot}\sim r_\star^{-3}$ instead of $P_{\rm mode}\sim r_\star^{-2}$).

\section{Hierarchy of Masses}

\label{sec:MassHierarchy}

We now investigate setups where perturbations theories is under control. One of the requirements for that is given in Eq.~\eqref{Cutoff}. In particular we see that in the limit where one mass is much bigger than the other, then
\ba
\frac{\Mred}{\Mtot}\ \xrightarrow{\ M_2\ll M_1\ }\  \frac{M_2}{M_1}\ll 1
\ea
and so we expect perturbation theory to be under control in that case. Taking the upper bound for $L$, $L=(\frac{\Mred }{\Mtot }v\Op \rstar )^{-1/4}$ the power radiated in that case is then given by
\begin{eqnarray}
\label{eq:power-with-mass-hierarchy}
\la P\ra 
  \approx  100\times\frac{\Mred }{\Mtot }\frac{v}{(\Op \rstar )^{3}}\Op ^{2}\,.
\end{eqnarray}
As a fiducial example of such a system, let us consider the Earth/Moon
system and use the power emission to estimate the rate of change of the orbit to check that we get a physically reasonable result.
In this case using $M_{\rm Earth}=5.97\times10^{24}{\rm kg}$, $M_{\rm Moon}=7.35\times10^{12}{\rm kg}$, $v/c=3.42\times10^{-6}$, and $\Op \rstar=648$ we find perturbation theory works with a cutoff $L=73$. Since the eccentricity of the earth-moon system is $\ep=0.05$ the approximation of zero eccentricity we have made in the calculation of the power is a good one. We find a radiated power
\begin{eqnarray}
\label{eq:power-earth-moon}
\la P\ra_{{\rm Moon}}  \approx 10^{-112}\Mpl^2\,,
\end{eqnarray}
which implies a rate of change of the orbit of
\begin{equation}
\label{eq:rbar-dot}
\dot{\rbar}=\frac{\rbar}{E_{\rm NR}}\frac{\rmd E_{\rm NR}}{\rmd t}
\end{equation}
where the system's non-relativistic energy is
\begin{equation}
\label{eq:enr}
E_{\rm NR}=\frac{1}{8 \pi^{2/3}}\frac{M_1 M_2}{\Mpl}\left(\frac{\Op^2}{\Mtot \Mpl}\right)^{1/3}\Mpl\,.
\end{equation}
Comparing this to the GR result for the power emission using the Peters-Mathews formula
(assuming $\ep=0$),
\be
P_{\rm Peters-Mathews}=\frac{32}{5}\frac{G^5}{c^5}\frac{M_{\bigoplus}^2 M_{\rm Moon}^2(M_{\bigoplus} + M_{\rm Moon})}{\bar{r}^5}\,,
\ee
we find
\be
\label{eq:rbar-dot-earth-moon-gal-vs-gr}
\frac{\dot{\bar{r}}_{gal}}{\dot{\bar{r}}_{GR}}\sim 10^{-33}\,,
\ee
which is utterly negligible ! We see that in that setup, the Vainshtein mechanism in the quartic Galileon is very active and prevents the scalar field from radiating almost any energy from the system.

\section{Hierarchy between Two Strong Coupling Scales}
\label{sec:SC_hierarchy}

A second situation worth mentioning where perturbation theory can remain under control in a binary system in the presence of quartic or quintic Galileon interactions is when these interactions do not dominate straight away but only far within the  strong coupling regime. To be more explicit we consider in what follows a hierarchy between the different strong coupling scales $\Lambda_3\ll \Lambda_4$ (or equivalently $\rstarFour\ll \rstarThree$)\footnote{Of course we still assume $\rstarFour\gg \bar r$, otherwise the quartic Galileon would not start dominating before one starts being strongly sensitive to the internal structure of extended object itself.}, so that as one probes shorter distances, the cubic Galileon starts dominating first and only very deep in the strong coupling region does the quartic Galileon take over.

To explore this situation, we construct the mode function in stages. We first approximate
$\hat{\bx}\phi$ as being equal to its leading order behavior
in each region. We start with the $\lag_{4}$ region, which
extends from the origin out to $\rstarFour$. We first need to determine
the correct boundary condition for the mode at the origin. It is not
necessarily correct to take the mode that is smooth at the origin,
because the equation of motion is singular there. So long as we are
are in the region $r\ll\Op ^{-1}$, the field can be approximated
as a power law, and we can extend the solution beyond the `crossover radii'
$\rstarThree $ and $\rstarFour$ by matching $\phi$ and its first derivative
at these radii. However once we reach the regime $r>\Op ^{-1}$,
the field begins to oscillate and we can no longer use this matching
procedure. Instead we can use the WKB approximation to extend the
solution out to infinity, where we can then fix the normalization.

It is clear that the details of this procedure depend on which strong
coupling region the scale $\Op ^{-1}$ falls into.
To get a sense of typical scales, take $\Lm_{3}=(1000{\rm km)^{-1}}$
and consider two solar mass binary pulsars with orbital period of
8 hours (which approximately describes the Hulse-Taylor pulsar), then
$\Op \rstarThree \sim10^{6}$. So the case where $\Op ^{-1}\ll \rstarThree $
is of most physical interest.
In what follows we will further assume that $\rstarFour \ll\Op ^{-1}\ll \rstarThree$, so that the crossing between slowly-varying and the WKB region occurs in region where the cubic Galileon dominates. In the case where $\rstarFour \gg\Op ^{-1}$, then the result is almost insensitive to the presence of the cubic Galileon.

\subsection{Mode functions}

The modes in the $\lag_{4}$ region are again $u_{\ell n}=\bar{u}_{\ell n}\sin(\omega_{\ell n}r)$, for $\ell>0$ (see eq.~\eqref{eq:multipole-norm}).
We then match this to modes in the $\lag_{3}$ region at $r=\rstarThree $
where the modes are given by \cite{deRham:2012fw}
\begin{equation}
\label{eq:modes-l3}
u_{\ell n}=a_{\ell n}\left(\frac{r}{\rstarThree }\right)^{1/4}J_{\nu_{\ell}}(\frac{\sqrt{3}}{2}\omega r)+b_{\ell n}\left(\frac{r}{\rstarThree }\right)^{1/4}Y_{\nu_{\ell}}(\frac{\sqrt{3}}{2}\omega r)\,,
\end{equation}
with $\nu_{\ell}=(2\ell+1)/4$. Matching $u_{\ell mn}$ and its first
derivative at $r=\rstarFour\ll\Op ^{-1}$, we find for  $\ell>0$ and to leading order in $\omega \rstarFour$
\begin{eqnarray}
\label{eq:modes-hierarchy-Lambda}
a_{\ell n} & = & \bar{u}_{\ell n}\frac{2+\ell}{1+2\ell}\Gamma(\nu_{\ell}+1)\left(\frac{\rstarThree }{\rstarFour}\right)^{1/4}\left(\frac{\sqrt{3}}{4}\right)^{-\nu_{\ell}}(n\Op \rstarFour)^{1-\nu_{\ell}}\\
b_{\ell n} & = & \mathcal{O}((n\Op \rstarFour)^{1+\nu_{\ell}})\,.
\end{eqnarray}
Now one can finally match this solution to the WKB regime in the limit  $\Op r\gg1$ to fix $\bar{u}_{\ell n}$.
Since $b_{\ell n}\approx0$, one can simply use the results derived in \cite{deRham:2012fw}
and set $a_{\ell n}=(9\pi/128\rstarThree ^{2})^{1/4}$,
so
\begin{eqnarray}
\label{eq:ubar-hierarchy-Lambda}
\bar{u}_{\ell n} & = & \frac{1+2\ell}{2+\ell}\left(\frac{9\pi}{128}\right)^{1/4}\left(\frac{\sqrt{3}}{4}\right)^{\nu_{\ell}}\frac{(n\Op \rstarFour)^{\nu_{\ell}-1}}{\Gamma(\nu_{\ell}+1)}\frac{\rstarFour^{1/4}}{\rstarThree ^{3/4}}\\
 & = & \beta_{\ell}(n\Op \rstarFour)^{\nu_{\ell}-1}\left(\frac{\rstarFour}{\rstarThree }\right)^{1/4}\frac{1}{\sqrt{\rstarThree }}\,,
\end{eqnarray}
where $\beta_{\ell}=\frac{1+2\ell}{2+\ell}\left(\frac{\sqrt{3}}{4}\right)^{\nu_{\ell}}\left(\frac{9\pi}{128}\right)^{1/4}\Gamma(\nu_{\ell}+1)^{-1}$
is a dimensionless prefactor. Note that $\beta_{\ell}$ is order
1 for $\ell=0$ but falls off rapidly with large $\ell$.

\subsection{Power}

Using these correctly normalized mode we find for a zero eccentricity source that
\ba
\label{eq:power-hierarchy-lambda}
\la P\ra&=&\frac{\pi}{3\Mpl^{2}}\frac{\Op }{\rstarThree }\sqrt{\frac{\rstarFour}{\rstarThree }}\sum_{n=0}^{\infty}\sum_{\ell m}n^{2\nu_{\ell}-1}\beta_{\ell}^{2}(\Op \rstarFour)^{2(\nu_{\ell}-1)}\\
&& \times\ \left|\frac{1}{\Tp }\int_{0}^{\Tp }\rmd t\ e^{-in\Op t}\int\rmd ^{3}x\ \sin(\omega_{\ell n}r) Y_{\ell m}(\theta,\varphi)\delta T(\mathbf{x},t)\right|^{2}\nn\,.
\ea
For $\ell\ll n\lcrit $ we have $\omega_{\ell n}\approx n\Op $.
Since $\Op \rbar \ll1$ we take the leading order behavior
$\sin(n\Op r)\approx n\Op r$, and find
\begin{eqnarray}
\label{eq:power-hierarchy-lambda-order-v1}
\la P\ra & = & \Op ^{2}\frac{\pi \Mred^{2}}{3\Mpl^{2}}\frac{\sqrt{\Op \rstarFour}}{\left(\Op \rstarThree \right)^{3/2}}v^{2}\sum_{n=0}^{\infty}\sum_{\ell m}n^{2\nu_{\ell}+1}\beta_{\ell}^{2}\left(\Op \rstarFour\right)^{2(\nu_{\ell}-1)}Y_{\ell m}(\frac{\pi}{2},0)^{2}\delta_{m,n}\,.
\end{eqnarray}
Consider the sum
\be
S(\ell)\equiv\sum_{n=0}^\infty n^{2\nu_\ell+1}\beta_\ell^2 (\Op\rstarFour)^{2(\nu_\ell-1)}Y_{\ell n}(\frac{\pi}{2},0)^2\,.
\ee
Here the small parameter $\Op \rstarFour$ plays the role of the velocity in the cubic Galileon by suppressing the higher order multipoles. At large $\ell$, using Stirling's approximation $\Gamma(z)\sim \sqrt{2\pi} z^{z-1/2}e^{-z}$ and approximating the sum over $n$ as an integral, we find the scaling behavior with $\ell$
\be
S(\ell)\sim \left(\frac{e\sqrt{3}}{2}\Op \rstarFour\right)^\ell \ell^{-1/2}\,.
\ee
Thus for $\Op \rstarFour < \sqrt{3} e/2\approx 2.3$ the summand becomes exponentially suppressed at large $\ell$. At very large $\Op \rstarFour$ the sum is dominated by the low multipole. Numerically we find that for $\Op \rstarFour=10^{-2}$ that $S(4)/S(2)\approx 10^{-3}$, and perturbation theory is then well under control as we recover a suppression at higher multipoles and moments.

\subsection{Quadrupole Radiation}

Focusing the power emitted by the Galileon quadrupole $\ell=2$, we have
\begin{equation}
\label{eq:dimensionless-constant}
\frac{\pi}{3}\beta_{2}^2\sum_{n=0}^{\infty}\sum_{m=-2}^{2}n^{7/2}Y_{2m}\(\frac{\pi}{2},0\)^{2}\delta_{m,n}
=\frac{\pi}{3}\beta_{2}^2\times2^{7/2}Y_{22}\(\frac{\pi}{2},0\)^{2}\approx 0.1\,,
\end{equation}
and so the power in the quadrupole is
\begin{equation}
\label{eq:power-quad-hierarchy-lambda}
\la P\ra_{{\rm \Lambda_3\ll \Lambda_4,}}^{(\ell=2)}=0.1\times\left(\frac{\Mred}{\Mpl}\right)^{2}\frac{(\Op \rstarFour)}{(\Op \rstarThree )^{3/2}}v^2\Op ^{2}\,.
\end{equation}
This result is almost identical to the cubic Galileon result in Eq.~\ref{eq:power-l3-quad}, but with one power of velocity replaced with $\Op \rstarFour$. The power is velocity enhanced relative to the pure $\lag_3$ case because the $\lag_4$ modes pick out the dipole moment of the source.

\section{Discussion}
\label{sec:discussion}

We find that the behavior of perturbations around spherical, time dependent backgrounds depends strongly on the presence of the fourth Galileon interaction. Unlike the case of the cubic Galileon considered in \cite{deRham:2012fw,Chu:2012kz} we find that studying perturbations around a naive static, spherically symmetric background is insufficient for computing the power emitted by a binary pulsar system. The key difference is the effective one dimensional metric seen by the fluctuations
\be\label{eq:eff-metric}
Z_{\mu\nu}\rmd x^\mu\rmd x^\nu\propto -\d t^2 + \d r^2 + \rstarFour^2 \d \Omega^2\,.
\ee
As a result of this one dimensional metric, the naive perturbation theory predicts that all multipole modes with fixed $n$ (defined by $\omega=n\Op$) contribute equally to the power until $\ell\sim n\Op r_\star$. Since for typical pulsar systems $\Op r_\star\sim 10^6$ this means a huge number of multipoles radiates with comparable strength. For simplicity we focused on systems with 0 orbital eccentricity $\ep=0$, but the divergence would be present even if we included the effects of eccentricity.

The resolution is that the `spherical background plus non-spherical perturbations' approximation breaks down. Physically we expect that this occurs because modes of arbitrarily high $\ell$ contribute to the power emitted, and higher $\ell$ modes are more sensitive to the lack of spherical symmetry in the system. This was checked by explicitly constructing the first and second order physical solutions to the equations of motion using the retarded propagator with a cutoff $L$ in the sum over multipoles, and taking their ratio at a point where their ratio was maximized. We found that whilst there is a range of parameters where perturbation theory could be trusted up to some $L$, for realistic pulsar systems we are forced to take $L<1$ so naive perturbation theory is never valid.

We expect that this result implies that there is an additional Vainshtein screening on top of the usual static and spherically symmetric screening. The breakdown of perturbation theory indicates that the perturbations themselves are nonlinear, and so contribute to their own Vainshtein suppression, in addition to the normal Vainshtein suppression from the background. As preliminary evidence in this direction, we note that the expression for the power with the cutoff in $L$ is more Vainshtein suppressed than the power calculated from any one mode.

We also studied two regimes where perturbation theory can be recovered, to check that the theory makes sensible (small power) predictions in these cases. The first situation is when there is a hierarchy of masses between the two bodies in the system (such as the Earth-Moon system). As a second example, we consider a hierarchy between the strong coupling scales of the cubic and the quartic Galileon.  The WKB oscillating behaviour then starts in the strong coupling region where the cubic Galileon dominates over the quartic one. We then find a suppression in the higher multipoles as expected from previous results. Nevertheless, as long as the interactions from the quartic Galileon are important on scales comparable to the size of the system $\bar r$, the system radiates as if it had a size $\rstarFour$ rather than $\bar r$.

We expect this result to be a generic feature of time dependent systems exhibiting the Vainshtein mechanism when the quartic Galileon is included. Thus the intuition gained from studying the static, spherically symmetric case may not directly apply to more complicated time-dependent (or less symmetric) systems. This two body system is surely the simplest generalization of the one body, static, spherically symmetric case and we already see at this level that a more detailed understanding is required.

Future work should involve going beyond this naive approximation to get a better analytic handle on the power emitted. We expect that one can find a background that takes into account the time dependent evolution.

\section*{Acknowledgments}

We would like to thank Yi-Zen~Chu and Mark~Trodden for useful discussions. AM is supported by the NSF GRFP program. A.J.T. was supported in part by the Department of Energy under grant DE-FG02-12ER41810.

\appendix

\section{Normalization of the Mode Functions}
\label{sec:App1}

In this section we check explicitly check the mode normalization by expanding around a regularized background.
Consider a smooth background source no longer localized at the origin, but rather spread over a radius $\varepsilon$,
\begin{equation}
\label{eq:app-background-source-regularized}
T_{0}^{\mu\nu}=\theta(\varepsilon -r)\frac{3M}{4\pi\varepsilon ^{3}}\delta_{0}^{\mu}\delta_{0}^{\nu}\,,
\end{equation}
where $\varepsilon \ll \rstarFour$. The background for $r<\varepsilon $ is
given by
\begin{equation}
\label{eq:app-background-pi-regularized}
\frac{E}{r}=\frac{\pi'(r)}{r}=\frac{\Lm_{4}^{2}}{\varepsilon }\left(\frac{M/\Mpl}{24\pi }\right)^{1/3}\,.
\end{equation}
The equation of motion for the fluctuations valid for $r<\varepsilon $
is then
\begin{equation}
\label{eq:app-fluctuation-eom-regularized}
\frac{6}{\Lm_{4}^{2}}\left(\frac{M/\Mpl}{24\pi }\right)^{2/3}\frac{1}{\varepsilon ^{2}}\left[-3\pdt^{2}\phi+\frac{1}{r^{2}}\pdr(r^{2}\pdr\phi)+\frac{1}{r^{2}}\pdO ^{2}\phi\right]=0\,,
\end{equation}
so that $Z_{rr}=Z_{\Omega\Omega}=\frac{1}{3}Z_{tt}=\frac{6}{\Lm_{4}^{2}}\left(\frac{M/\Mpl}{24\pi }\right)^{2/3}\frac{1}{\varepsilon ^{2}}$.
Now there is no singularity at $r=0$, so modes inside $r<\varepsilon$ satisfy
\begin{equation}
\label{eq:app-modes-inside-regularized}
u_{\ell mn}^{<}(r)=\bar{u}_{\ell mn}j_{\ell}(\sqrt{3}\omega r)\approx\bar{u}_{\ell mn}(\sqrt{3}\omega r)^{\ell}\,.
\end{equation}
Outside the source,  for $r>\varepsilon $  the modes  are given by
\begin{equation}
\label{eq:app-modes-outside-regularized}
u_{\ell mn}^{>}=A\cos\omega_{\ell}r+B\sin\omega_{\ell}r\approx A+B\omega r\,,
\end{equation}
using the approximations $\ell\ll\Op \rstarFour$ and $r\ll \rstarFour$.

Now the equations of motion are smooth at $r=\varepsilon $, and so the
appropriate matching conditions are $u_{\ell mn}^{>}(\varepsilon )=u_{\ell mn}^{<}(\varepsilon )$
and $\pdr u_{\ell mn}^{>}(\varepsilon )=\pdr u_{\ell mn}^{<}(\varepsilon )$.
Solving these for $A$ and $B$ and taking the ratio, we find
\begin{equation}
\label{eq:app-norm-solution-regularized}
\frac{A}{B}=\frac{\ell-1}{\ell}\omega\varepsilon\,,
\end{equation}
so in the limit $\varepsilon \rightarrow0$ the term going as $B$ dominates
over the constant going as $A$ (this is also valid for the dipole $\ell=1$).

For the monopole ($\ell=0$)  on the other hand the contributions are different
and in that case the same analysis shows that the constant term $A$ dominates. This is reason why the mode normalization differs for $\ell=0$ and $\ell>0$ as can be seen in eqs.~\eqref{eq:monopole-norm} and \eqref{eq:multipole-norm}.

\section{Small Strong Coupling Scale for the Quintic Galileon in $\lag_3$ Region}\label{sec:App2}

Here we find that $Z_{rr}$ and $Z_{\Omega\Omega}$ are dominated by the
pieces that are independent of $\Lm_{4}$ and $\Lm_{5}$,
and so these reproduce the equations of motion for fluctuations from
the cubic Galileon. There is a contribution from $\lag_{5}$
to $K_{t}$ however:
\ba
\label{eq:app-Ztt-lag3-small-lag5}
Z_{tt}(r)&=&-\frac{32}{\sqrt{3}\pi^{3/2}(1/3)^{3/2}}\left(\frac{\rstarThree }{r}\right)^{9/2}\left(\frac{\Lm_{3}}{\Lm_{5}}\right)^{9/2}+6\sqrt{\frac{1}{9\pi}}\left(\frac{\rstarThree }{r}\right)^{3/2}\\
&& + {\rm \ higher\ corrections}\nn\,.
\ea
The higher corrections piece refers to terms that are higher order
in $r/\rstarThree $ and also terms that are suppressed by powers of $\Lm_{3}/\Lm_{4}$
(which must be a small ratio for this region to even exist) which
do not dominate for $r>\rstarFour$.

The first term here is comparable to the second term for $r\sim(\Lm_{4}^{4}/\Lm_{3}\Lm_{5}^{3})\rstarFour$.
Thus if we take $\Lm_{5}\geq(\Lm_{4}/\Lm_{3})^{1/3}\Lm_{4}$
then we can ignore this term in the $\lag_{3}$ region.

If we do take a small $\Lm_{5}$ then we find that the equation
of motion has the form of Laplace's equation in 3D for $r<(\Lm_{4}^{4}/\Lm_{3}\Lm_{5}^{3})$.
\begin{equation}
\label{eq:app-eom-lag3-small-lag5}
\hat{\bx}\phi=2\sqrt{\frac{1}{9\pi}}\left(\frac{\rstarThree }{r^{2}}\right)^{3/2}\left(4\pdr^{2}\phi+\frac{2}{r}\pdr\phi+\frac{1}{r^{2}}\left(-\frac{16}{\pi (1/3)^{2}}\rstarThree ^{2}\left(\frac{\Lm_{3}}{\Lm_{5}}\right)^{9/2}\pdt^{2}\phi+\pdO^{2}\phi\right)\right)\,.
\end{equation}
This is similar to what occurs in the $\lag_{4}$ region, where
time and angular derivatives appear without a relative factor of $1/r^{2}$.
However a detailed analysis of this situation is beyond the scope of this work.

\section{Sum over the Multipoles and Moments}\label{sec:App3}

We start with the sum in \eqref{eq:regularized-sum}
\begin{equation}
\label{eq:app-regularized-sum-zero-eccentricty}
S_{0}(\nmax)=\sum_{n=0}^{\nmax}\sum_{\ell=1}^{n\lcrit }\sum_{m=-\ell}^{\ell}n^{2}Y_{\ell m}\left(\frac{\pi}{2},0\right)^{2}\cos^{2}\left(\frac{m\pi}{2}\right)\delta_{n,m}\,.
\end{equation}
Now focusing on multipoles with $\ell\gg n$ (for a realistic pulsar system, $\Op \rstar \sim10^{6}\gg1),$
\begin{equation}
\label{eq:app-Ylm_approx}
Y_{\ell m}\(\frac{\pi}{2},0\)\approx\frac{1}{\pi}\cos\left((\ell+m)\frac{\pi}{2}\right)\,.
\end{equation}
Note that $Y_{\ell m}\(\frac{\pi}{2},0\)^{2}\cos^{2}\left(\frac{m\pi}{2}\right)
\approx\frac{1}{\pi^{2}}\cos^{2}((\ell+m)\frac{\pi}{2})\cos^{2}(\frac{m\pi}{2})=\frac{1}{\pi^{2}}\cos^{2}(\frac{\ell\pi}{2})\cos^{2}(\frac{m\pi}{2})$.
Then the sum becomes
\begin{equation}
\label{eq:app-regularized-sum-zero-eccentricty-y-approx}
S_{0}(\nmax)\approx\frac{1}{\pi^{2}}\sum_{n=0}^{\nmax}\sum_{\ell=1}^{n\lcrit }\sum_{m=-\ell}^{\ell}n^{2}\cos^{2}\left(\frac{\ell\pi}{2}\right)\cos^{2}\left(\frac{m\pi}{2}\right)\delta_{n,m}\,.
\end{equation}
Now we can figure out the value of the two inner sums as a function
of $n$. Doing these sums amounts to counting the number of nonzero
terms, because the summand is either 1 or 0. Explicitly
\begin{eqnarray}
\label{eq:app-inner-sum}
s(n) & \equiv & n^{2}\sum_{\ell=1}^{n \lcrit}\sum_{m=-\ell}^{\ell}\cos^{2}\left(\frac{\ell\pi}{2}\right)\cos^{2}\left(\frac{m\pi}{2}\right)\delta_{n,m}\\
 & = & n^{2}\left(\frac{n\lcrit}{2}\right)\cos^{2}\left(\frac{n\pi}{2}\right)\,.
\end{eqnarray}
Then we arrive at a final approximation for the sum
\begin{eqnarray}
\label{eq:app-regularized-sum-solved}
S_{0}(\nmax) \approx \frac{\lcrit }{2\pi^{2}}\sum_{n=0}^{\nmax}n^{3}\cos^{2}(\frac{n\pi}{2}) \approx  \frac{1}{16\pi^{2}}\lcrit \nmax^{4}\,.
\end{eqnarray}

\section{First and Second order field fluctuations}\label{sec:App4}
\subsection{Retarded Propagator}

We normalize the retarded propagator so that
\be\label{eq:app-def-retarded-prop}
\hat{\square}G_{R}(x,x')=\delta^{4}(x-x')\,,
\ee
where $\hat{\square}$ is given in \eqref{dAlembertian} (strictly speaking this normalization procedure is only acceptable as long as the modes are continued all the way to the oscillating WKB regime).
In this procedure the mode functions are real, $u_{\ell\omega}(r)=u_{\ell\omega}^{*}(r)$ and so we have (using the fact that $\sum_m Y_{\ell m}Y^*_{\ell m}$ is real)
\be\label{eq:app-ret-g-sum-after-im}
G_{R}(x,x')=-\theta(t-t')\int \d\omega\sin\omega(t-t')\sum_{\ell=0}^{\infty}\sum_{m=-\ell}^{\ell}u_{\ell\omega}(r)u_{\ell\omega}(r')Y_{\ell m}(\theta,\varphi)Y_{\ell m}^{*}(\theta',\varphi')\,.
\ee
Then given a source $J(x)$ ($\Box \phi= J$), the field is given by
\ba\label{eq:app-field-from-ret-prop}
\phi(x) & = & \int \d^{4}x'G_{R}(x,x')J(x')\\
 & = & -\int \d^{4}x'\int \d\omega\theta(t-t')\sin\omega(t-t')\\
 &&\times\sum_{\ell=0}^{\infty}\sum_{m=-\ell}^{\ell}u_{\ell\omega}(r)u_{\ell\omega}(r')Y_{\ell m}(\theta,\varphi)Y_{\ell m}^{*}(\theta',\varphi')J(x')\nn\,.
\ea
Now define
\be\label{eq:app-def-jlm}
J_{\omega\ell m}(t')=\int \d^{3}x'u_{\ell\omega}(r')Y_{\ell m}^{*}(\theta',\varphi')J(\vec{x}',t')\,,
\ee
and since the source is periodic we can write
\be\label{app-fourier-jlm}
J_{\omega\ell m}(t)=\sum_{n=-\infty}^{\infty}f_{n}(\omega,\ell,m)e^{in\Omega_{P}t}\,.
\ee
Then the expression for the field is given by
\be\label{eq:app-phi-soln}
\phi(x)=-\int \d t'\int \d\omega\theta(t-t')\sin\omega(t-t')\sum_{n=-\infty}^{\infty}\sum_{\ell m}u_{\ell\omega}(r)Y_{\ell m}(\theta,\phi)f_{n}(\omega,\ell,m)e^{in\Omega_{P}t'}\,.
\ee
Performing the integral over $t'$ yields
\ba\label{eq:app-def-phiR-phiS}
\phi(x) & = & \sum_{n=-\infty}^{\infty}\sum_{\ell m}\int \d\omega\ e^{in\Omega_{P}t}f_{n}(\omega,\ell,m) u_{\ell\omega}(r)Y_{\ell m}(\theta,\varphi)\\
&&\times\left[-\frac{i\pi}{2}\left(\delta(\omega-n\Omega_{P})-\delta(\omega+n\Omega_{P})\right)+\frac{\omega}{\omega^{2}-n^{2}\Omega_{P}^{2}}\right]\nn\\
 & = & \phi_{R}(x)+\phi_{S}(x)\,,
\ea
where $\phi_{R}(x)$ is the piece with the delta functions and $\phi_{S}(x)$
is the piece without the delta functions. As we will see, $\phi_{R}$
is the radiating piece of the solution and dominate in the WKB regime while $\phi_{S}$ is essentially
the static piece and is negligible in the WKB regime and does not contribute to the radiated power.

We first focus on simplifying $\phi_R$. The $\omega$ integrals are trivial because of the delta functions.
Carrying through the simplification we find
\be\label{eq:app-phiR-solved}
\phi_{R}(x)=\pi\sum_{n=0}^{\infty}\sum_{\ell m}u_{\ell n}(r)Y_{\ell m}(\theta,\varphi)\int\frac{\d t'}{T_{P}}\int \d^{3}x'\sin n\Omega_{P}(t-t')Y^*_{\ell m}(\theta',\varphi')u_{\ell n}(r')J(\vec{x}',t')\,.
\ee
Note that for $\Omega_{P}=0$, $\phi_{R}=0$. This part of the field
is only present for time dependent sources, and it is responsible
for radiation. We could have seen this above too, we had $\delta(\omega-n\Omega_{P})-\delta(\omega+n\Omega_{P})=0$
for $\Omega_{P}=0$.

As for the second part of the field $\phi_S$,
\begin{eqnarray}
\label{eq:app-phi-S-solved}
\phi_{S}(x) & = & \sum_{n=-\infty}^{\infty}\sum_{\ell m}\int \d\omega e^{in\Omega_{P}t}u_{\omega\ell}Y_{\ell m}(\theta,\phi)f_{n}(\omega,\ell,m)\frac{\omega}{\omega^{2}-(n\Omega_{P})^{2}}\\
 & = & 2\sum_{n=0}^{\infty}\sum_{\ell m}\int\frac{\d t'}{T_{P}}\int \d^{3}x'\int \d\omega\frac{\omega}{\omega^{2}-(n\Omega_{P})^{2}}Y_{\ell m}(\theta,\phi)Y_{\ell m}^{*}(\theta',\phi')\\
&&\times u_{\omega\ell}(r)u_{\omega\ell}(r')\cos n\Omega_{P}(t-t')J(\vec{x}',t')\nn\,.
\end{eqnarray}
Note that this contribution does not vanish for $\Op=0$, so it includes the contribution
from the static propagator.

This term is suppressed by $1/n\Op$ compared to $\phi_R$ as can be seen by the following argument.
We do not expect the $n=0$ mode to contribute to the power because
the static piece is not oscillating so derivatives acting on it are
suppressed compared to the oscillating parts of the field. Note that
for large $n$ the integral over $\omega$ is basically 0 because
$u_{\omega\ell}(r)u_{\omega\ell}(r')$ is even in $\omega$ (at least
for all of the kinds of modes we have considered) and $\omega/(\omega^{2}-\omega_{0}^{2})$
is odd in $\omega$. For large $n$, the pole is far enough away from
0 that the bounds on the integral can be approximated as $\int_{-\infty}^{\infty}$,
which gives 0 because the integrand is odd.

So since $n=0$ and large $n$ modes are small compared to the radiation
piece, we expect that this term should be small
in the radiation zone.
We can explicitly see how this works for the $\lag_4$ modes by using the integrals
\begin{eqnarray}\label{eq:app-omega-integrals}
\int_{0}^{\infty}\d x\frac{\sin(ax)\sin(bx)}{p^{2}-x^{2}}  =  -\frac{\pi}{2p}
\left\{\begin{array}{lcl}
\cos(ap)\sin(bp) &\ \ {\rm if}\ \ & a>b>0\\
\frac 12 \sin(2ap)&\ \ {\rm if}\ \ & a=b>0\\
\sin(ap)\cos(bp)&\ \ {\rm if}\ \ & b>a>0
 \end{array}\right.\,.
\end{eqnarray}
If we use the modes appropriate for $\lag_4$ for $0<\ell\ll\lcrit$, that is $u_n=\frac{1}{\sqrt{n\Op}r_\star}\sin(n\Op r)$, then the integral over $\omega$ in \ref{eq:app-phi-S-solved} is of this form with $x=\omega$, $a=r$, $b=r'$, $p=n\Op$. Then we can see explicitly that for $\Op r \gg 1$ that $\phi_S$ is suppressed with respect to $\phi_R$ by $1/n\Op$.

\subsection{Field in the WKB regime}

Now we explicitly construct $\phi^{(1)}$ in the WKB regime. We
consider only the radiating part, $\phi_{R}^{(1)}$ (and drop
the subscript). The source for $\phi^{(1)}$ is $J^{(1)}(x)=-\frac{1}{\sqrt{6}M_{{\rm Pl}}}\delta T(x)$. Here we plug in the modes
appropriate for $\mathcal{L}_{4}$.
\begin{eqnarray}\label{eq:app-phi1-with-deltaT-source}
\phi^{(1)}(x) & = & \pi\sum_{n=0}^{\infty}\sum_{\ell m}u_{\ell n}(r)Y_{\ell m}(\theta,\varphi)\\
&&\times\int\frac{\d t'}{T_{P}}\int \d^{3}x'\sin n\Omega_{P}(t-t')Y^*_{\ell m}(\theta',\varphi')u_{\ell n}(r')J(\vec{x}',t')\nn\\
 & = & \frac{\pi}{\sqrt{6}M_{{\rm Pl}}r_{*}^{2}}\sum_{n\ell m}\sum_{j=1,2}M_{j}\bar{r}_{j}(1+(-1)^{jm})\sin n\Omega_{P}rY_{\ell m}(\theta,\varphi)Y_{\ell m}(\frac{\pi}{2},0)\\
 &&\times\int\frac{\d t'}{T_{P}}\theta(\ell-n\ell_{{\rm crit}})\sin n\Omega_{P}(t-t')e^{-im\Omega_{P}t}f_{\epsilon}(t')\nn\,,
\end{eqnarray}
where
\be\label{eq:app-def-bars}
\bar{r}_{1,2}=\frac{M_{2,1}}{\Mtot}\bar{r}\,.
\ee
We define the reduced mass
\be\label{eq:app-def-m-red}
\sum_{j}M_{j}\bar{r}_{j}=\frac{M_{1}M_{2}}{\Mtot}\bar{r}+\frac{M_{2}M_{1}}{\Mtot}\bar{r}=2\mathcal{M}\bar{r}\,,
\ee
where $\mathcal{M}$ is the reduced mass. Considering a source with no eccentricity, we have $f_{0}(t')=1$.

Then the integral over $t'$ simplifies to
\be\label{eq:app-t-int}
\int\frac{\d t'}{T_{P}}\sin n\Omega_{P}(t-t')e^{-im\Omega_{P}t}=\frac{i}{2}\left(e^{-in\Omega_{P}t}\delta_{n,m}-e^{in\Omega_{P}t}\delta_{n,-m}\right)\,.
\ee
We can simplify the resulting expression for $\phi$ (remember $n\geq0$ and use the Shortley phase rule $Y_{\ell,-m}(\theta,\varphi)=(-1)^{m}Y_{\ell m}^{*}(\theta,\varphi)$, and the facts $Y_{\ell m}(\theta,0)\in\mathbf{R}$
and $Y_{\ell m}(\theta,\varphi)=Y_{\ell m}(\theta,0)e^{im\varphi}$),
\ba
&&\sum_{m=-\ell}^{\ell}Y_{\ell m}(\theta,\varphi)Y_{\ell m}(\frac{\pi}{2},0)\left(e^{-in\Omega_{P}t}\delta_{n,m}-e^{in\Omega_{P}t}\delta_{n,-m}\right)\nn \\
&&\hspace{20pt} = -2i\sum_{m=0}^{\ell}Y_{\ell m}(\theta,0)Y_{\ell m}\(\frac{\pi}{2},0\)\sin(n\Omega_{P}t-m\varphi)\delta_{n,m}\,.
 \label{eq:app-lsum}
\ea
Then we are left with
\ba
\label{eq:app-phi1-solved}
\phi^{(1)}(x)&=&\frac{2\pi\mathcal{M}\bar{r}}{\sqrt{6}M_{{\rm Pl}}r_{*}^{2}}\sum_{\ell=0}^{\infty}\sum_{m=0}^{\ell}\theta(\ell-m\ell_{{\rm crit}})\cos^{2}\frac{m\pi}{2}Y_{\ell m}(\theta,0)Y_{\ell m}\(\frac{\pi}{2},0\)\\
&&\times \sin m\Omega_{P}r\sin(m\Omega_{P}t-m\varphi)\delta_{n,m}\nn\,.
\ea
We now consider a regulated $\phi^{(1)}$. We
truncate the sum over $\ell$ at the cutoff $L<\ell_{\rm crit}$. Since
we are in the 0 eccentricity case this will imply a cutoff in $N$
because the kronecker delta forces $n\leq\ell$.
\ba
\label{eq:app-phi1-solved-with-L-cutoff}
\phi^{(1)}(x,L)&=&\frac{2\pi\mathcal{M}\bar{r}}{\sqrt{6}M_{{\rm Pl}}r_{*}^{2}}\sum_{\ell=0}^{L}\sum_{m=0}^{\ell}\cos^{2}\frac{m\pi}{2}Y_{\ell m}(\theta,0)Y_{\ell m}\(\frac{\pi}{2},0\)\\
&&\times\sin m\Omega_{P}r\sin(m\Omega_{P}t-m\varphi)\,.\nn
\ea
We are now interested in finding the maximum value of $\phi^{(1)}$.
This amounts to picking $\theta=\pi/2,\varphi=0$ and then considering points on the radial lightcone
$r=t$ with the specific values $r=t=(2k+1)\pi\Op/2$, so that $\sin m\Omega_{P}r\times\sin m\Omega_{P}t=1$ for any even  $m$.
Call this set of parameters $x_{\rm max}$ (note
that $x_{\rm max}$ is not a unique set of parameters, it is just any choice
that satisfies these conditions). In this case (using the approximation for $Y_{\ell m}\(\pi/2,0\)$)
\be\label{eq:app-phi1-with-L-cutoff-at-xmax}
\phi^{(1)}(x_{\rm max},L)\approx   \frac 1{4.5}\frac{2\pi\mathcal{M}\bar{r}}{\sqrt{6}M_{{\rm Pl}}r_{*}^{2}}\frac{L^{2}}{\pi^{2}}= \frac{4\mathcal{M}\bar{r}}{9\pi\sqrt{6}M_{{\rm Pl}}r_{*}^{2}}L^{2}\,.
\ee
The factor of $1/4.5$ is an approximation. Naively one would guess
this factor is 1/4 because we are taking roughly half of the $m$
terms and half of the $\ell$ terms. Numerically it seems that $4.5$
for this factor is a better fit.

\subsection{Second Order Fluctuations}

For $r>\Omega_{P}^{-1}$ we ignore the $\phi_{S}$ piece compared
to the $\phi_{R}$ piece, since we are interested in the derivatives
of the field.
We  first compute explicitly the source for  $\phi^{(2)}$,
\begin{eqnarray}
\label{source}
J^{(2)}(x) & = & \frac{3}{\Lambda^{6}}\Big[\square\pi_{0}\left(\square\phi^{(1)}\right)^{2}-\square\pi_{0}\left(\nabla_{\mu}\nabla_{\nu}\phi^{(1)}\right)^{2}\\
&&-2\nabla_{\mu}\nabla_{\nu}\pi_{0}\nabla^{\mu}\nabla^{\nu}\phi^{(1)}\square\phi^{(1)}
+2\nabla_{\mu}\nabla^{\nu}\pi_{0}\nabla_{\nu}\nabla^{\lambda}\phi^{(1)}\nabla_{\lambda}\nabla^{\mu}\phi^{(1)}\Big]\nn\\
 & + & \frac{1}{\Lambda^{3}}\left[\left(\square\phi^{(1)}\right)^{2}-\left(\nabla_{\mu}\nabla_{\nu}\phi^{(1)}\right)^{2}\right]\nn\,,
\end{eqnarray}
where the covariant derivatives are taken with respect to the flat 3d metric in spherical coordinates.
As discussed earlier, since we work in the WKB regime we can do an expansion of derivatives in powers of $1/\Op$, $\pd^2 f\sim\left((n\Op)^2+\frac{n\Op}{r}f+\cdots\right)f$.
Thus to leading order in $\Op$,
\be\label{eq:app-j2-leading-is-zero}
J^{(2)}(x)=-\frac{6}{(24\pi c_{4})^{1/3}}\left(\frac{M_{{\rm tot}}}{M_{{\rm Pl}}}\right)^{1/3}\frac{1}{\Lambda^{4}r}\left(-\left(\partial_{t}^{2}\phi^{(1)}\right)^{2}
+\left(\partial_{r}^{2}\phi^{(1)}\right)^{2}\right)=0\,,
\ee
and one needs to work to next to leading order in $\Op$ to see any relevant contributions. In this case the terms in the line of \eqref{source} that actually arose from $\mathcal{L}_3$ are the dominant ones,
\begin{eqnarray}
\label{eq:app-j2-solved}
J^{(2)} & = & -\frac{1}{\Lambda^{3}}\left[\left(\partial_{r}^{2}\phi^{(1)}\right)^{2}
-2\left(\partial_{t}\partial_{r}\phi^{(1)}\right)^{2}+\left(\partial_{t}^{2}\phi^{(1)}\right)^{2}\right]\\
 & = & \frac{2}{\Lambda^{3}}\left[\left(\partial_{t}\partial_{r}\phi^{(1)}\right)^{2}-\left(\partial_{r}^{2}\phi^{(1)}\right)^{2}\right]\,,
\end{eqnarray}
where we have used the fact that $\pd_r \phiFirstOrder=\pd_t \phiFirstOrder$ to first order in $\Op^{-1}$.
For fixed $\ell,\ell',m,m'$ this looks like
\ba\label{eq:def-jllmm}
J_{\ell\ell'mm'}^{(2)} & = & \frac{2}{\Lambda^{3}}\left(\frac{2\pi\mathcal{M}\bar{r}}{\sqrt{6}M_{{\rm Pl}}r_{*}^{2}}\right)^{2}\Omega_{P}^{4}m^{2}m^{'2}\\
&\times&\cos^{2}\frac{m\pi}{2}Y_{\ell m}(\theta,0)Y_{\ell m}(\frac{\pi}{2},0)Y_{\ell'm'}(\theta,0)Y_{\ell'm'}\(\frac{\pi}{2},0\)\nn\\
 & \times & \big(\cos m\Omega_{P}r\cos m'\Omega_{P}r\cos(m\Omega_{P}t-m\varphi)\cos(m'\Omega_{P}t-m'\varphi)\nn\\
 &&-\sin m\Omega_{P}r\sin m'\Omega_{P}r\sin(m\Omega_{P}t-m\varphi)\sin(m'\Omega_{P}t-m'\varphi)\big)\nn\,.
\ea
Knowing the source, the equation of motion for the second order field fluctuation $\phi^{(2)}(x)$ is
\be\label{eq:app-eom-phi2}
\hat{\square}\phi^{(2)}=\left(\frac{r_{*}}{r}\right)^{2}
\left(-\partial_{t}^{2}+\partial_{r}^{2}+\frac{1}{r_{*}^{2}}\partial_{\Omega}^{2}\right)\phi^{(2)}(x)=J^{(2)}(x)\,.
\ee
In the WKB region $r>\Omega_{P}^{-1}$ we expect the solution
to take the from
\be\label{eq:app-phi2-wkb}
\phi^{(2)}\sim A(r,\Omega_{S})\sin(\omega r+\varphi_{r})\sin(\omega t+\varphi_{t})\,,
\ee
with $A(r,\Omega_{S})$ a slowly varying function. Considering the worst case scenario, we can neglect the
angular dependence. In order to match with the source we take $A(r)=ar^{2}$ leading to the ansatz
\ba\label{eq:phi2-ansatz}
\phi^{(2)}(x) & = & \left(\frac{r}{r_{*}}\right)^{2}\sum_{\ell=0}^{L}\sum_{\ell'=0}^{L}\sum_{m=0}^{\ell}\sum_{m=0}^{\ell}a_{\ell\ell'mm'}Y_{\ell m}(\theta,0)Y_{\ell'm'}(\theta,0)\\
&&\times\left(\sin m\Omega_{P}r\sin m'\Omega_{P}r\sin(m\Omega_{P}t-m\varphi)\sin(m'\Omega_{P}t-m\varphi)\right)\nn\,.
\ea
To satisfy the equations of motion we must set
\be\label{eq:allmm-solved}
a_{\ell\ell'mm'}=\frac{2}{\Lambda^{3}}\left(\frac{2\pi\mathcal{M}\bar{r}}{\sqrt{6}M_{{\rm Pl}}r_{*}^{2}}\right)^{2}\Omega_{P}^{2}mm^{'}\cos^{2}\frac{m\pi}{2}Y_{\ell m}\(\frac{\pi}{2},0\)\,.
\ee
So now we have an expression for $\phi^{(2)}$ valid in the regime
where $\mathcal{L}_{3}$ terms dominate.
Now evaluating the second order field fluctuation at the set of point $x_{\rm max}$ where the first order fluctuations take their maximal value, we get
\be\label{eq:app-phi2-with-L-cutoff}
\phi^{(2)}(x_{\rm max},L)=\left(\frac{r}{r_{*}}\right)^{2}\frac{2}{\Lambda^{3}}\left(\frac{2\pi\mathcal{M}\bar{r}}{\sqrt{6}M_{{\rm Pl}}r_{*}^{2}}\right)^{2}\Omega_{P}^{2}\left(\sum_{\ell=0}^{L}\sum_{m=0}^{\ell}mY_{\ell m}(\frac{\pi}{2},0)^{2}\cos^{2}\frac{m\pi}{2}\right)^{2}\,,
\ee
with the sum being approximated by
\be\label{eq:app-L-sum}
\sum_{\ell=0}^{L}\sum_{m=0}^{\ell}mY_{\ell m}(\frac{\pi}{2},0)^{2}\cos^{2}\frac{m\pi}{2}\approx\frac{L^{3}}{4\pi^{2}}\,,
\ee
so the second order field fluctuation at these set of points is
\be\label{eq:app-phi2-with-L-cutoff-solved}
\phi^{(2)}(x_{\rm max},L)=\left(\frac{r}{r_{*}}\right)^{2}\frac{2}{\Lambda^{3}}\left(\frac{2\pi\mathcal{M}\bar{r}}{\sqrt{6}M_{{\rm Pl}}r_{*}^{2}}\right)^{2}\Omega_{P}^{2}\frac{L^{6}}{16\pi^{4}}\,.
\ee
As we see in section \ref{sec:Power_1}, (see eq.~\eqref{eq:phi12-ratio-solved}), this usually implies a breaking of perturbation theory in many physical situations.

\section{Validity of Perturbation Theory in the Cubic Galileon}
\label{sec:App5}

Finally we give a simple check on the validity of perturbation theory in the case of the cubic Galileon. Unlike in the quartic and quintic case, we find that the velocity parameter $v$ acts as a small parameter which ensures the convergence of the perturbative expansion.
The properly normalized modes found in Ref.~\cite{deRham:2012fw} are given by
\be
u_{\ell \omega}(r)=\(\frac{9\pi}{128}\)^{1/4}\frac{1}{\sqrt{\rstar}}\(\frac{r}{\rstar}\)^{1/4}J_{\nu_{\ell}}\(\frac{\sqrt{3}}{2}\omega r\)\, ,
\ee
with
\ba
\nu_{\ell}=\left\{
\begin{array}{ccl}
(2\l+1)/4 & {\rm for} &  \ \ \l>0  \\
-1/4 & {\rm for} &\ \ \l=0
\end{array}
\right. \,.
\ea
We compute $\phi_R$ in the WKB regime using the retarded propagator, as in appendix~\ref{sec:App4}. We find that for $\ell>0$
\be
\phi_R(\vec{x},t)=-\(\frac{\sqrt{3}\pi}{128}\)^{1/2}\sum_{\ell=0}^\infty \sum_{m=-\ell}^\ell \sum_{n=0}^\infty a_{n\ell m}(t) Y_{\ell m}(\theta,\varphi)\(\frac{r}{\rstar}\)^{1/4} \frac{\cos\(\frac{\sqrt{3}}{2} n \Op r + P\)}{\sqrt{n \Op r}}\, ,
\ee
where $P$ is an irrelevant phase factor. The time dependent coefficients $a_{n\ell m}(t)$ are given by\\
$\bullet$ \ \ \ \textbf{$\l>0$ modes}
\be
a_{n\ell m}(t)=\frac{M_{\ell , m}}{\Mpl}\frac{\alpha_\ell}{\rstar}n^{\nu_\ell} \(\frac{\rbar}{\rstar}\)^{1/4}  v^{\nu_\ell}\times\int_0^{T_p}\frac{\rmd t'}{T_P}\sin\(n\Op(t-t')\)e^{-i m t'} f_\epsilon(t')\, ,
\ee
with $M_{\ell, m}\equiv M_1 (M_2/\Mtot)^{(\ell+1)/2}+(-1)^m M_2 (M_1/\Mtot)^{(\ell+1)/2}$, and  $\alpha_\ell\equiv Y_{\ell m}(\frac{\pi}{2},0)(\frac{\sqrt{3}}{4})^{\nu_\ell}\Gamma\(\nu_\ell+1\)^{-1}$, and $f_\epsilon(t)\equiv (1-\epsilon^2)(1+\epsilon \cos(\Op t))^{-1}$and by\\
$\bullet$ \ \ \ \textbf{$\l=0$ modes}
\be
a_{n00}(t)=\frac{\mathcal{M}}{\Mpl}\frac{\alpha'_0}{\rstar}n^{7/4} \(\frac{\rbar}{\rstar}\)^{1/4} v^{7/4}\times\int_0^{T_P}\frac{\rmd t'}{T_P}\sin\(n\Op(t-t')\) f_\epsilon^2(t')\, ,
\ee
where $\mathcal{M}$ is the reduced mass and where $\alpha'_0\equiv Y_{00}(\frac{\pi}{2},0)2^{-(7/4)}\Gamma(7/4)^{-1}$. The monopole and dipole must be treated separately from the other multipoles because at leading order in $v$ the $\ell=0$ and $1$ part of $\phi_R$ are zero by energy and momentum  conservation. Thus the monopole and dipole both receive an extra power of $v^2$ suppression compared to the scaling $v^{\nu_{\ell}}$ from the other multipoles.

Now we check the in the case of the cubic Galileon, the action for the perturbations $\phi$ contains, in addition to
\eqref{eq:lag-quad}, the cubic term
\be
S^{(3)}_{\phi}=-\int \d^4 x\frac{1}{3\sqrt{6}\Lambda^3}(\partial \phi) ^2 \Box \phi\,,
\ee
which should be small in order for perturbation theory to be under control.
To demonstrate
this is consistent we consider the effect of these terms in the two
distinct regions $r \ge \Op^{-1}$ and $r \le \Op^{-1}$. In the
region $r \ge \Op^{-1}$ but $r \le r_*(=\rstarThree)$, a simple estimate of the
ratio of the cubic to quadratic terms in the action gives, for $\ell>1$
\be
\frac{S^{(3)}_{\phi}}{S^{(2)}_{\phi}}\sim
\(\frac{r}{r_*}\)^{3/2}\frac{\omega^2 \phi_R(r)}{\Lambda^3}\sim (\omega \rbar)^{\nu_{\ell}+1/4} (\omega
r)^ {5/4}\,,
\ee
which is small for all $\ell>1$ if we take $r\sim \Op^{-1}$. For the monopole and dipole we get
\ba
\left.\frac{S^{(3)}_{\phi}}{S^{(2)}_{\phi}}\right|_{\ell=0} \sim  v^2 \hspace{20pt}{\rm and}\hspace{20pt}
\left.\frac{S^{(3)}_{\phi}}{S^{(2)}_{\phi}}\right|_{\ell=1} \sim  v^3\,,
\ea
again evaluated at $r\sim \Op^{-1}$ in terms of the velocity $v=\Omega_P \rbar$.
We see that the extra $v^2$ suppression for the monopole and dipole is crucial because it means that the ratio is also small when $\ell=0,1$.
Thus if we
evaluate the flux on a sphere of radius $r \sim \Omega_P^{-1}$
where linear theory is valid.

\bibliographystyle{jhep}
\bibliography{refs}

\end{document}